\documentclass[oldversion]{aa}
\pdfoutput=1

\usepackage[intlimits]{amsmath}
\usepackage{txfonts}
\usepackage[american]{babel}
\usepackage{graphicx}
\usepackage{xspace}

\usepackage{natbib}
\newcommand{\simm}{\ensuremath{\mathord{\sim}}}
\newcommand{\Nabla}{\ensuremath{\vec{\nabla}}}
\newcommand{\abs}[1]{\ensuremath{\left|#1\right|}}

\newcommand{\degree}{\ensuremath{^\circ}}
\newcommand{\unit}[1]{\ensuremath{\,\mathrm{#1}}}
\newcommand{\MURaM}{\texttt{MURaM}\xspace}
\newcommand{\mean}[1]{\langle#1\rangle}
\newcommand{\lcr}{\ensuremath{\lambda_\mathrm{cr}}}
\newcommand{\lci}{\ensuremath{\lambda_\mathrm{ci}}}
\newcommand{\lr}{\ensuremath{\lambda_\mathrm{r}}}
\newcommand{\taus}{\ensuremath{\tau_\mathrm{s}}}

\newcommand{\imagetop}[1]{\vtop{\null\hbox{#1}}}

\begin{document}

\title{Vortices in simulations of solar surface convection}
\author{R. Moll \and  R. H. Cameron\ \and M. Sch{\"u}ssler}
\institute{Max-Planck-Institut f\"{u}r Sonnensystemforschung, Max-Planck-Stra{\ss}e 2, 37191 Katlenburg-Lindau, Germany}
\date{Accepted on August 1, 2011}
\abstract{We report on the occurrence of small-scale vortices in simulations of
the convective solar surface.  Using an eigenanalysis of the velocity gradient
tensor, we find the subset of high-vorticity regions in which the plasma is
swirling.  The swirling regions form an unsteady, tangled network of filaments
in the turbulent downflow lanes.  Near-surface vertical vortices are underdense
and cause a local depression of the optical surface. They are potentially
observable as bright points in the dark intergranular lanes. Vortex features
typically exist for a few minutes, during which they are moved and twisted by
the motion of the ambient plasma. The bigger vortices found in the simulations
are possibly, but not necessarily, related to observations of granular-scale
spiraling pathlines in ``cork animations'' or feature tracking.}
\keywords{Magnetohydrodynamics (MHD) -- Sun: photosphere -- Sun: granulation}
\maketitle

\section{Introduction}

Solar surface convection is characterized by upflows of hot plasma in Mm-scale
granules and downflows of radiatively cooled plasma in a network of relatively
narrow intergranular lanes \citep[for a review, see][]{2009Nordlund}.  The
turbulent nature of the downflows, together with angular momentum conservation,
suggests the formation of vortical flows at scales on the order of the width of
the downflow lanes or smaller. In fact, vortex structures regularly appear in
simulations of solar surface convection with or without magnetic fields
\citep[e.g.,][]{1998Stein,2004Voegler,2010Muthsam,2011Kitiashvili,2011Shelyag}.

There is observational evidence for photospheric whirl flows on larger scales
between $5\unit{Mm}$ and $20\unit{Mm}$ \citep{1988Brandt,1997Simon,2009Attie}.
Granular-scale vortical motions associated with intergranular downflows (``sink
holes'') were found only more recently on the basis of high-resolution
observations \citep{2008Bonet,2010Bonet,2011Dominguez}.  In such studies, the
horizontal velocities are inferred either by local correlation tracking or by
direct tracing of individual features, such as magnetic bright points
\citep[e.g.,][]{1996Berger,2010Balmaceda}.  \citet{2010Steiner} found dark
lanes moving into granules in high-resolution images obtained with the
balloon-borne telescope SUNRISE. By comparison with numerical simulations, they
identified these lanes as horizontally oriented vortex tubes, presumably
originating from the overturning motion at the boundaries of the intergranular
lanes \citep[cf.][]{1998Stein}.  Small-scale swirl events were also reported in
chromospheric observations \citep{2009Wedemeyer}; their relation to surface
convection is unclear \citep[cf.][]{2010Carlsson}.

In this paper, we report on the detection of sub-intergranular scale vortices
in simulations. We begin with a description of our simulations and the
technique used for vortex identification in Sect.~\ref{sec:methods}.  In
Sect.~\ref{sec:results}, we present a statistical analysis of vortices in our
simulations, study the physical properties of selected specimens, and examine
possible observational signatures. We summarize our findings in
Sect.~\ref{sec:discussion}.

\begin{figure*}[t]
\includegraphics[width=\linewidth]{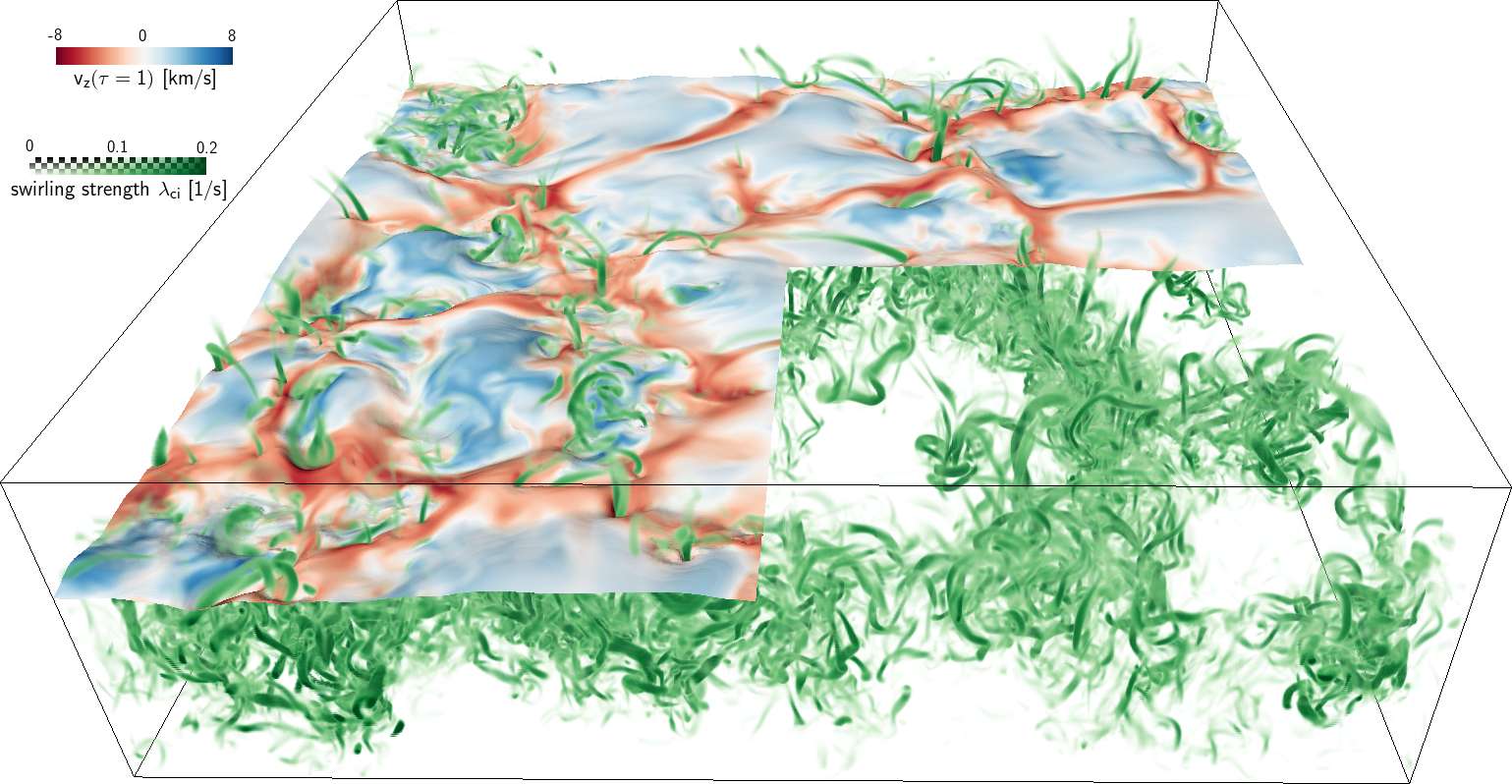}
\caption{Snapshot of the swirling strength (green volume rendering) and the
optical surface color-coded with vertical velocity (downflows in red and
upflows in blue) in Run~C.  The size of the box shown is
\(4.8\times4.8\times1.4\unit{Mm^3}\). The optical surface is hidden in the
lower right quadrant, uncovering the swirling structure in the subsurface
layers.}
\label{fig:Im3D}
\end{figure*}

\section{Methods}
\label{sec:methods}

\subsection{Simulations of solar convection}

\begin{table}[b]
\caption{Parameters of the simulations. The vertical range is given with
respect to the height at which the horizontal mean temperature is
\(\mathord{\approx}7000\unit{K}\).}
\centering
\begin{tabular}{@{}cccc@{}}
\hline\hline
Run & Box size [$\mathrm{Mm}^3$]  & Cell size [$\mathrm{km}^3$] & Vertical range [km] \\
\hline
    C    &  \(4.86\times4.86\times1.4\)  & \(7.5\times7.5\times10\) & \(-890<z<510\) \\
    D    &  \(12\times12\times6.144\)    & \(20.83\times20.83\times16\) & \(-5560<z<584\) \\
    H    &  \(4.86\times4.86\times1.4\)  & \(4\times4\times4\) & \(-884<z<516\) \\
\hline
\end{tabular}
\label{tab:parameters}
\end{table}

We used the \MURaM code \citep{2003Voegler,2005Voegler} to solve the MHD
equations in the context of solar surface convection, including radiative
energy transfer and an equation of state which incorporates the effects of
partial ionization.  The simulations are performed on a Cartesian grid, using a
4th order central difference scheme for the spatial discretization and a
short-characteristics scheme for the radiative transfer.  At the bottom of the
domain, open boundary conditions allow for the in- and outflow of matter:
\(\partial_z \vec{v}=0\) for outflows and \(v_{x,y} = \partial_z v_z = 0\) for
inflows, i.e., inflows are vertical.  The pressure and the entropy of inflowing
matter are chosen such that the total mass in the domain is approximately
constant and the radiative energy loss is consistent with the solar luminosity.
The top boundary is closed and the horizontal boundaries are periodic.  The
magnetic field is assumed to be vertical at the top and bottom boundaries:
\(B_{x,y}=\partial_z B_z=0\).  For the radiative energy transfer, a gray
approximation is assumed. The equation of state being used is appropriate for
the conditions in the solar photosphere and the near-surface convection zone.

The parameters of the simulations considered here are listed in
Table~\ref{tab:parameters}.  Runs~C and~H cover the same vertical range at
intermediate and high resolutions, respectively.  Both simulations include
weak, initially random magnetic fields and have previously been used to study
near-surface dynamo action \citep{2007Voegler,2010Graham,2011Moll}.  Their high
resolution allows us to study the fine details of small vortices.  Run~D does
not include a magnetic field and uses a lower resolution, allowing us to probe
the convection zone at greater depths.

\subsection{Vortex identification}
\label{sec:vortex}

Despite there being a clear intuitive notion of what a vortex is (viz., the
rotation of fluid parcels about a possibly moving axis), it is surprisingly
difficult to come up with a formal definition. In fact, contemporary research
in fluid mechanics is still concerned with finding unambiguous vortex
identification schemes \citep[e.g.,][]{2004Haller,2007Kolar}.  Given its
frequent use in the literature, it should be noted that a high vorticity
$\omega = \abs{\Nabla \times \vec{v}}$ is not a sufficient indicator for the
presence of a vortex, because $\omega$ can also be high in shear flows without
rotation.

To detect vortices, we use the so-called ``swirling strength'', or $\lci$,
criterion \citep{1999Zhou}, which is based on an eigenanalysis of the velocity
gradient tensor $U_{ij}\coloneqq(\partial_j v_i)$. Vortices are defined to be
regions where $U$ has a pair of complex conjugate eigenvalues.  A large
unsigned imaginary part $\lci$ implies a region of ``strong swirling'', i.e.,
(part of) a vortex.  In the case of rigid rotation, the value of $\taus
\coloneqq 2\pi/\lci$ gives the revolution period of the rotating flow.  Unlike
many other methods for vortex detection, the $\lci$ criterion is applicable
also in the case of compressible hydrodynamics \citep{2009Kolar}.

Depending on the signs of the real eigenvalues $\lr$ and the real part of the
complex eigenvalues $\lcr$, we discriminate four types of vortices
\citep{1999Haimes}:
\begin{enumerate}
\item $\lr < 0$, $\lcr < 0$: spiraling inward, converging
\item $\lr > 0$, $\lcr < 0$: spiraling inward, diverging
\item $\lr < 0$, $\lcr > 0$: spiraling outward, converging
\item $\lr > 0$, $\lcr > 0$: spiraling outward, diverging
\end{enumerate}
Here, converging and diverging refer to the fluid motion along the direction of
the eigenvector corresponding to $\lr$, which is identified as the direction of
the vortex. Note that this direction is not necessarily orthogonal to the
``swirling plane'' spanned by the real and imaginary parts of the complex
eigenvectors.  We distinguish vortices by their inclination angle $\iota$ with
respect to the vertical direction.

The identification of vortices via $\lci$ is Galilean invariant, because it
only relies on derivatives of the velocity field.  Streamlines, however, are
not invariant under Galilean transformations.  Hence the streamlines in a
particular frame of reference (e.g., the one which is at rest with respect to
the computational box) do not always show circular orbits where vortices are
detected.  A strong vortex, however, is expected to produce vortical
streamlines for a large range of observer velocities.

Although the definition of the swirling strength is purely local (as is the
definition of vorticity), it can be shown that for a volume of rigidly rotating
fluid, the swirling strength is uniform over the entire interior of the volume.

\section{Results}
\label{sec:results}

\subsection{Occurrence of vortices}

\begin{figure*}[t]
\centering
\begin{tabular}{r@{\hskip10pt}l}
\includegraphics[height=.4\linewidth]{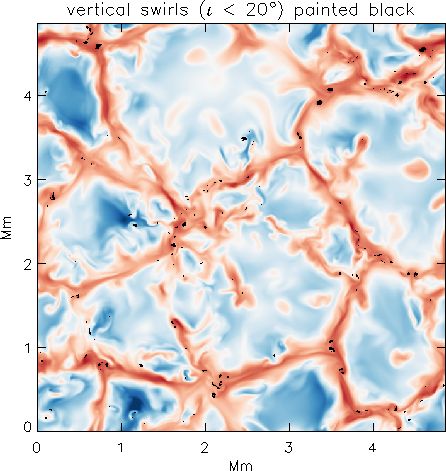} &
\includegraphics[height=.4\linewidth]{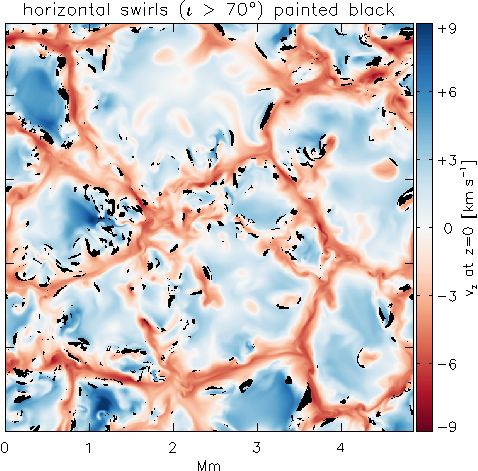}
\end{tabular}
\caption{Vertical velocities at the average height of the optical surface
(\(T\approx7000\unit{K}\)).  Black spots indicate vortices with a swirling
period \(\taus < 120\unit{s}\) and an inclination of less than $20\degree$ with
respect to the vertical (horizontal) direction in the \textit{left-hand (right
hand)} panel.}
\label{fig:mapvor}
\end{figure*}

\begin{figure}[t]
\includegraphics[width=\linewidth]{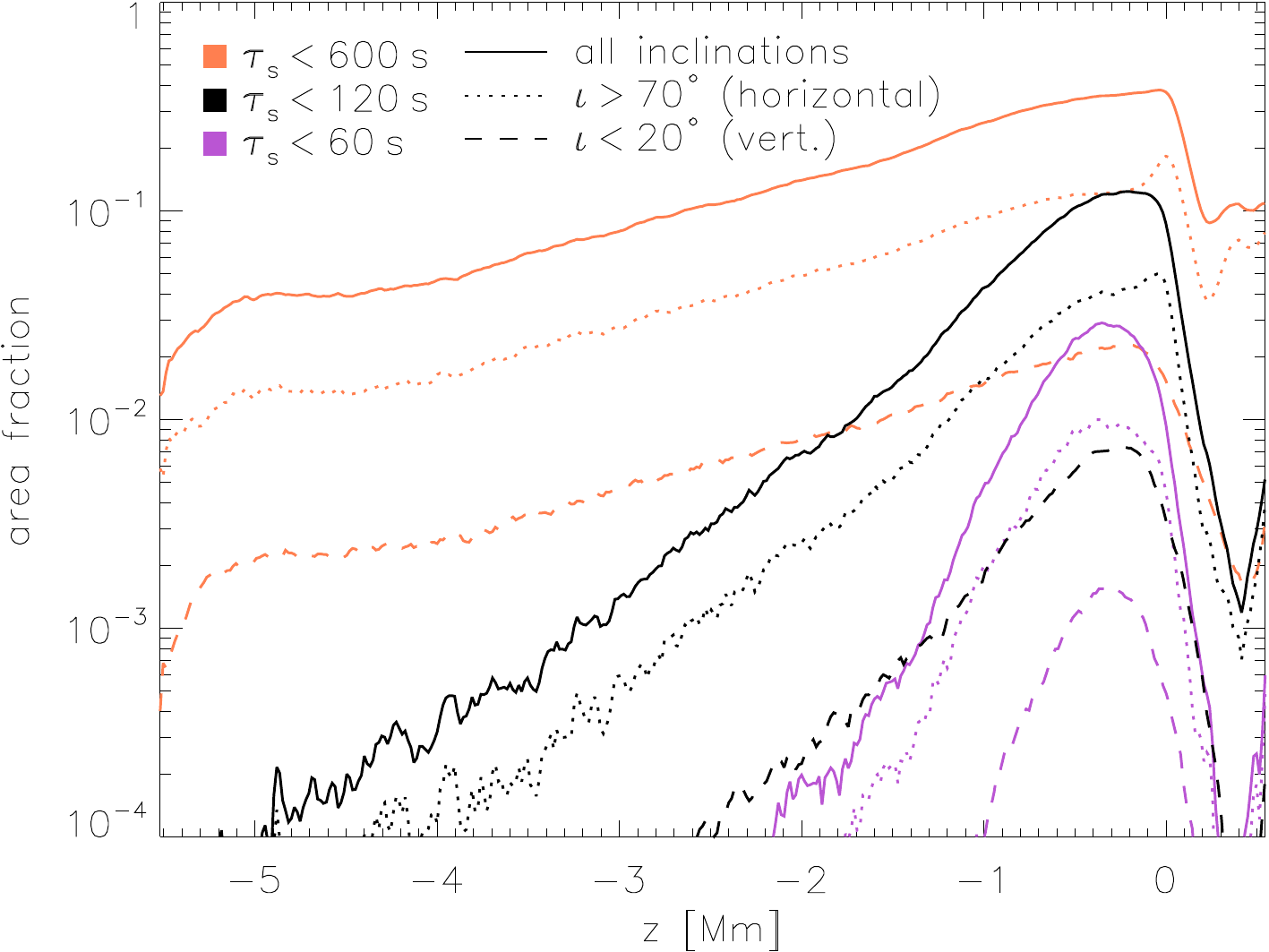}
\caption{Horizontal area fraction of grid cells with swirling periods below a
given threshold $\taus$ as a function of height in Run~D.  \(z=0\) corresponds
to the average height of the optical surface.}
\label{fig:depthcov}
\end{figure}

\newcommand{\myscale}{.51}
\begin{figure*}[p]
\centering
\begin{tabular}{r@{\hskip20pt}l}
    \includegraphics[scale=\myscale]{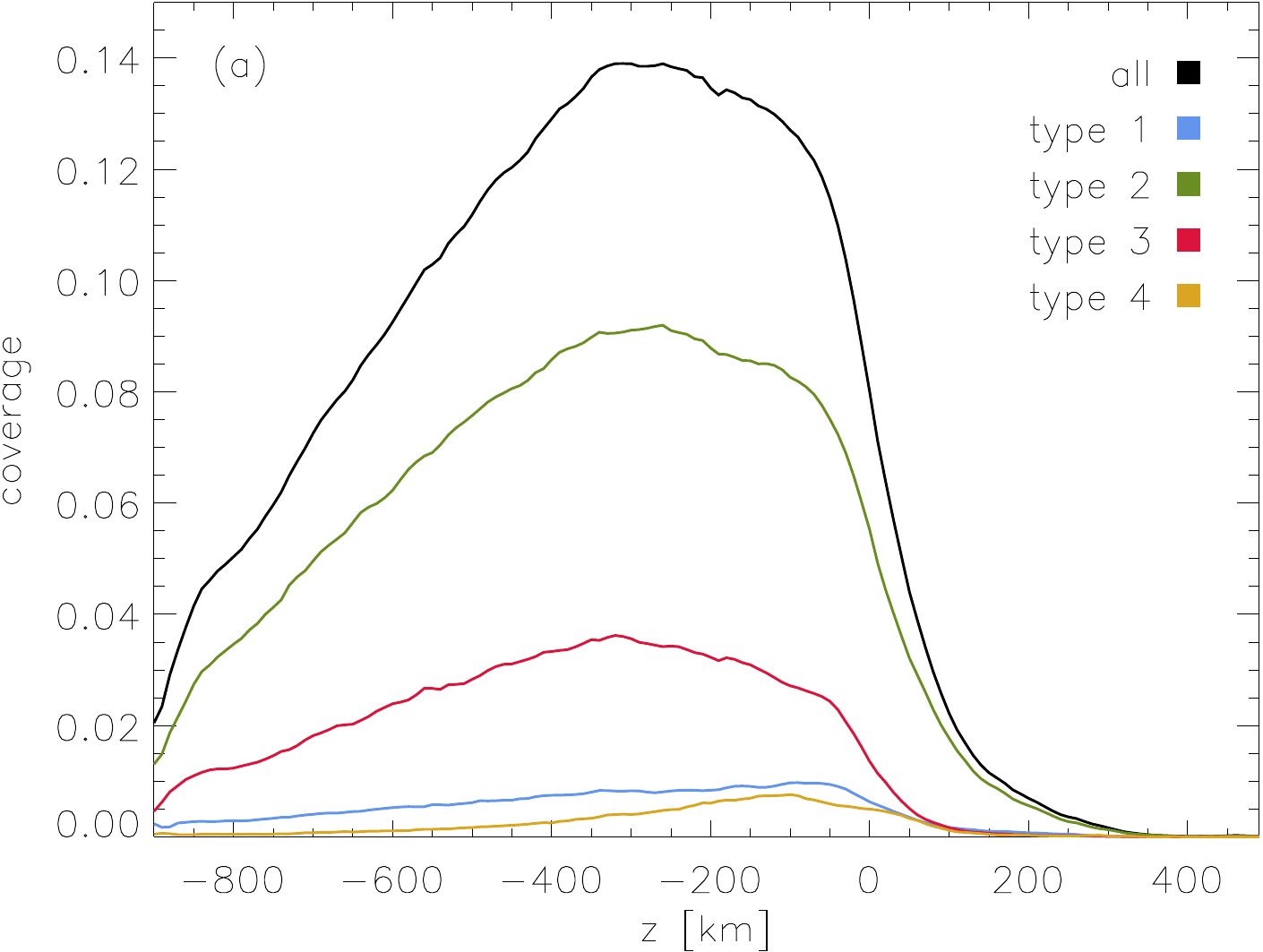} &
    \includegraphics[scale=\myscale]{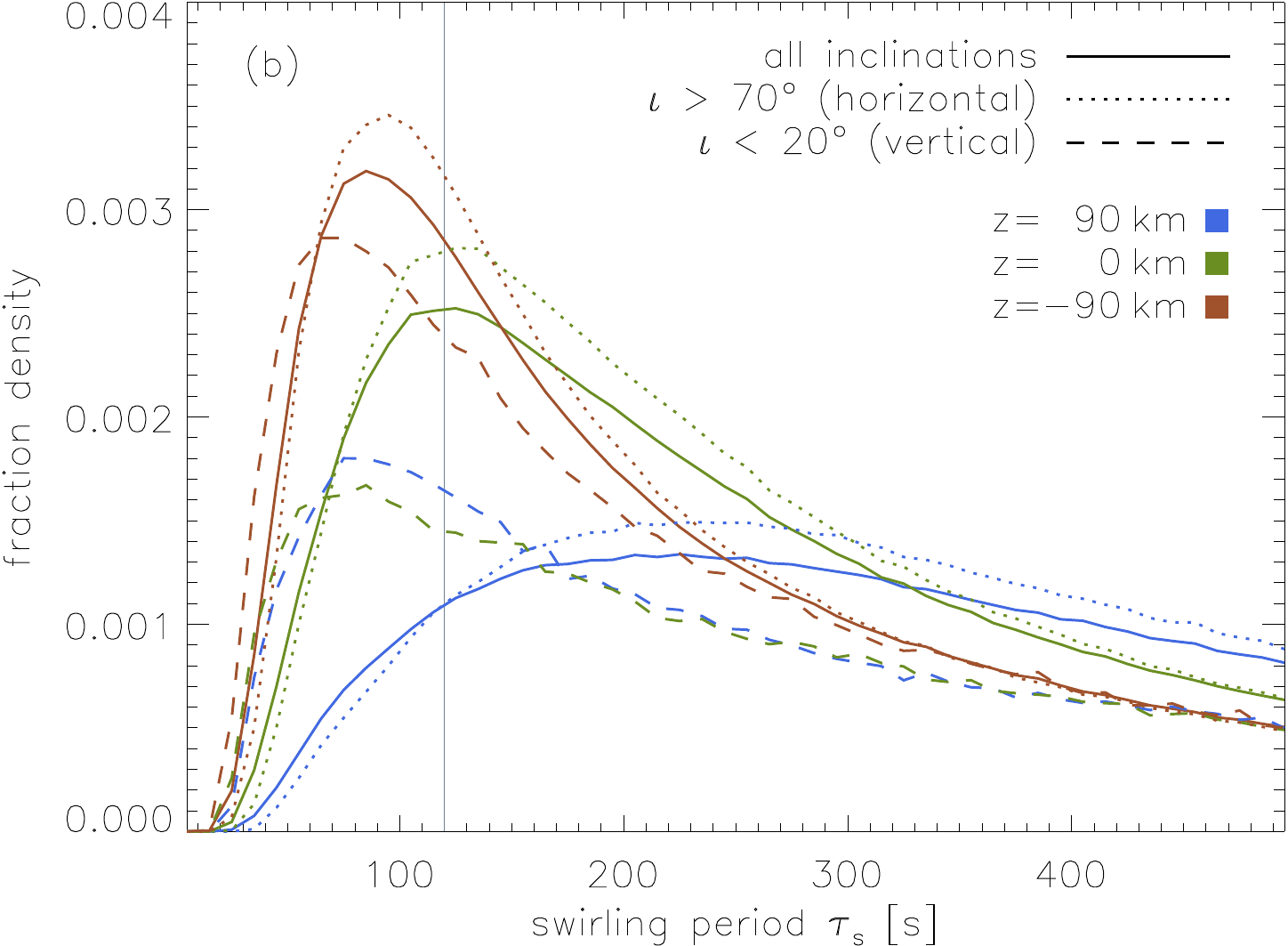} \\
    \includegraphics[scale=\myscale]{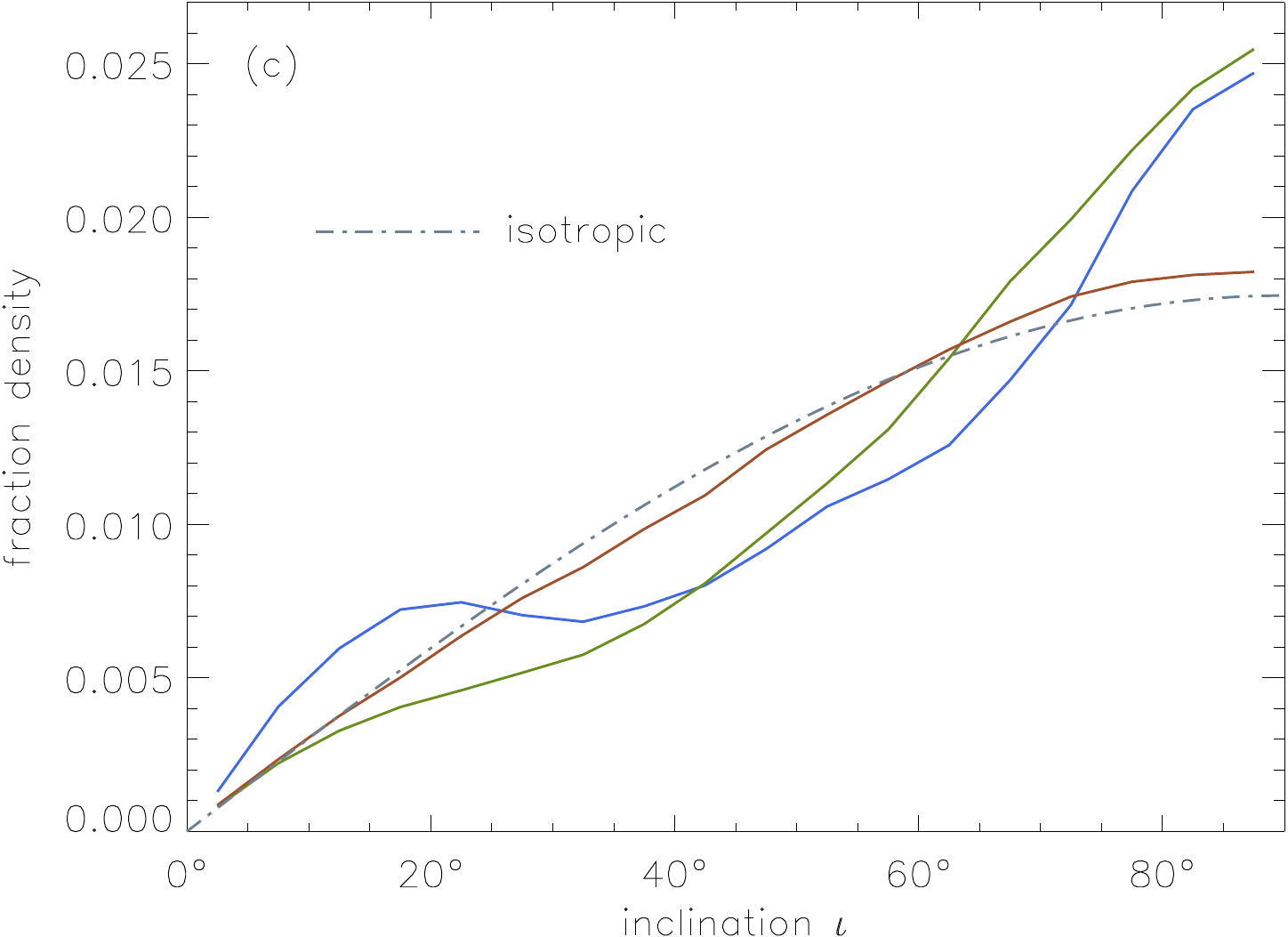} &
    \includegraphics[scale=\myscale]{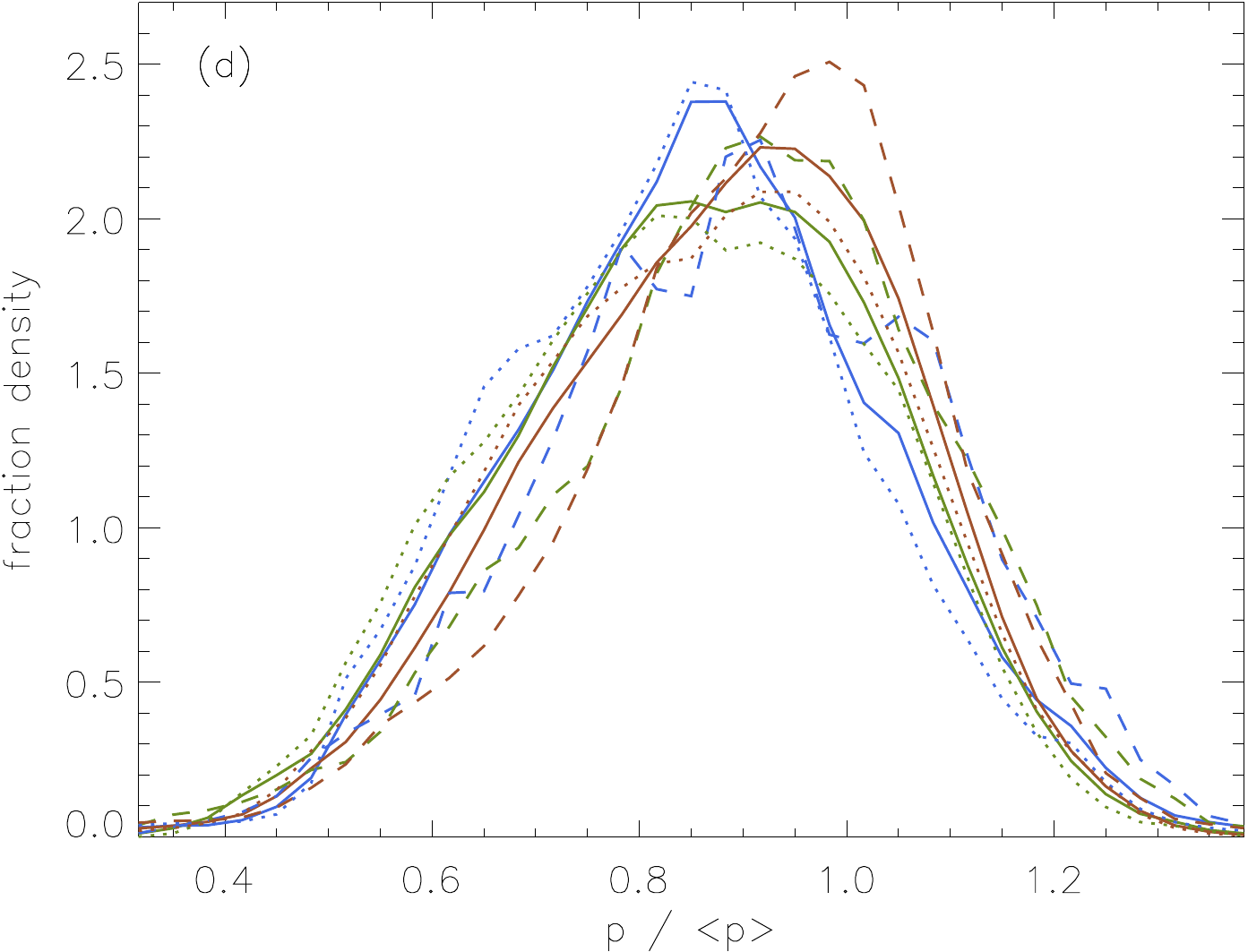} \\
    \includegraphics[scale=\myscale]{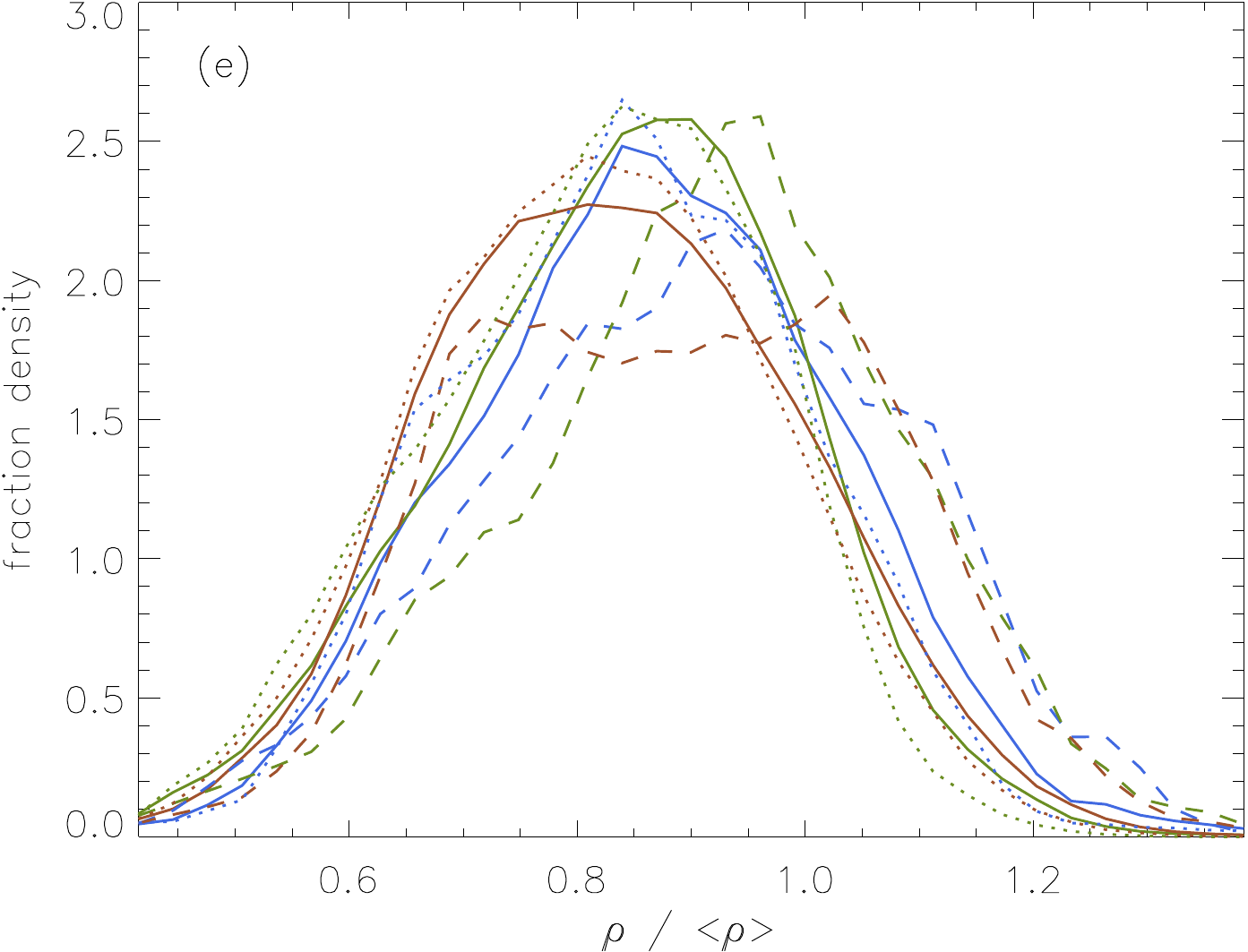} &
    \includegraphics[scale=\myscale]{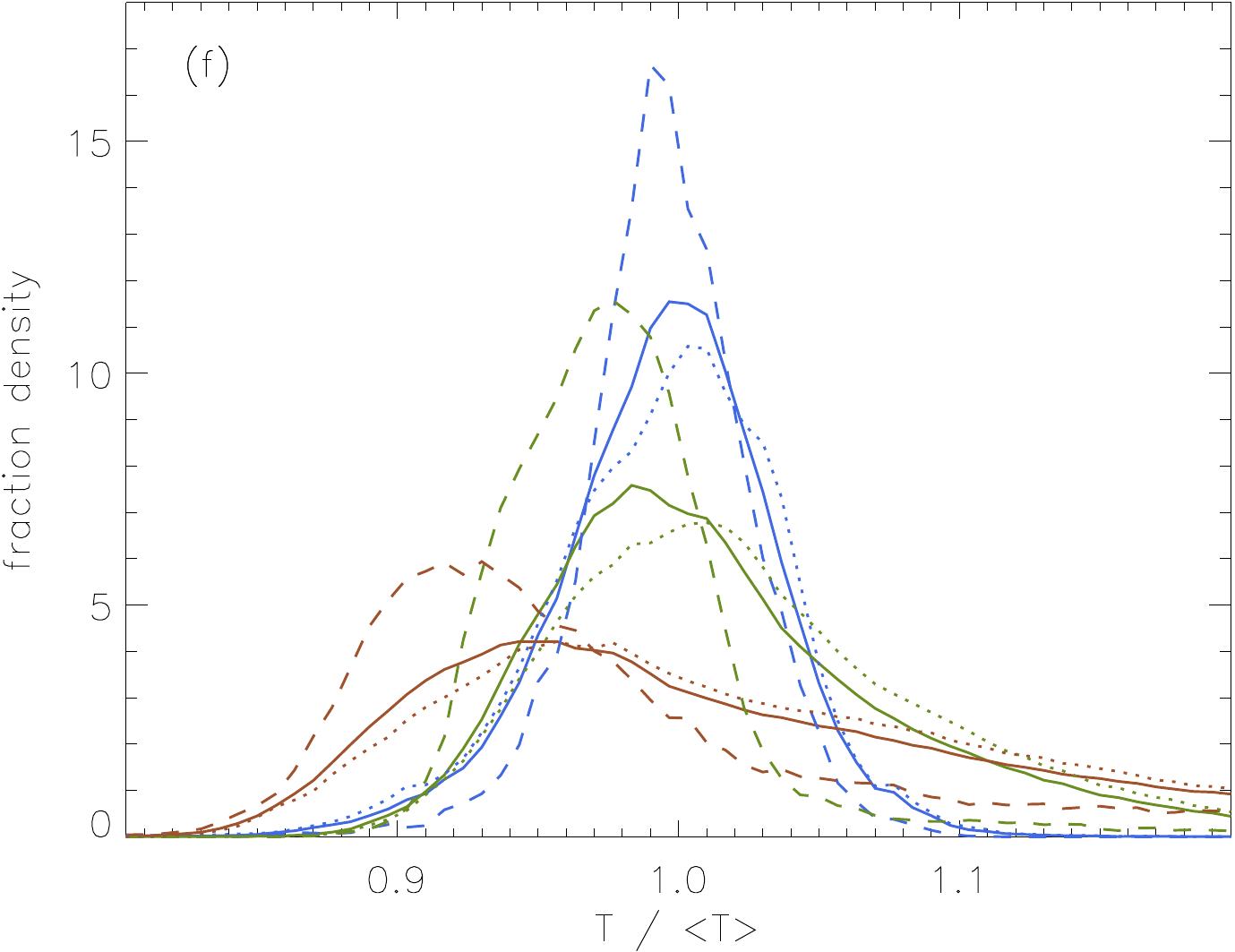} \\
    \includegraphics[scale=\myscale]{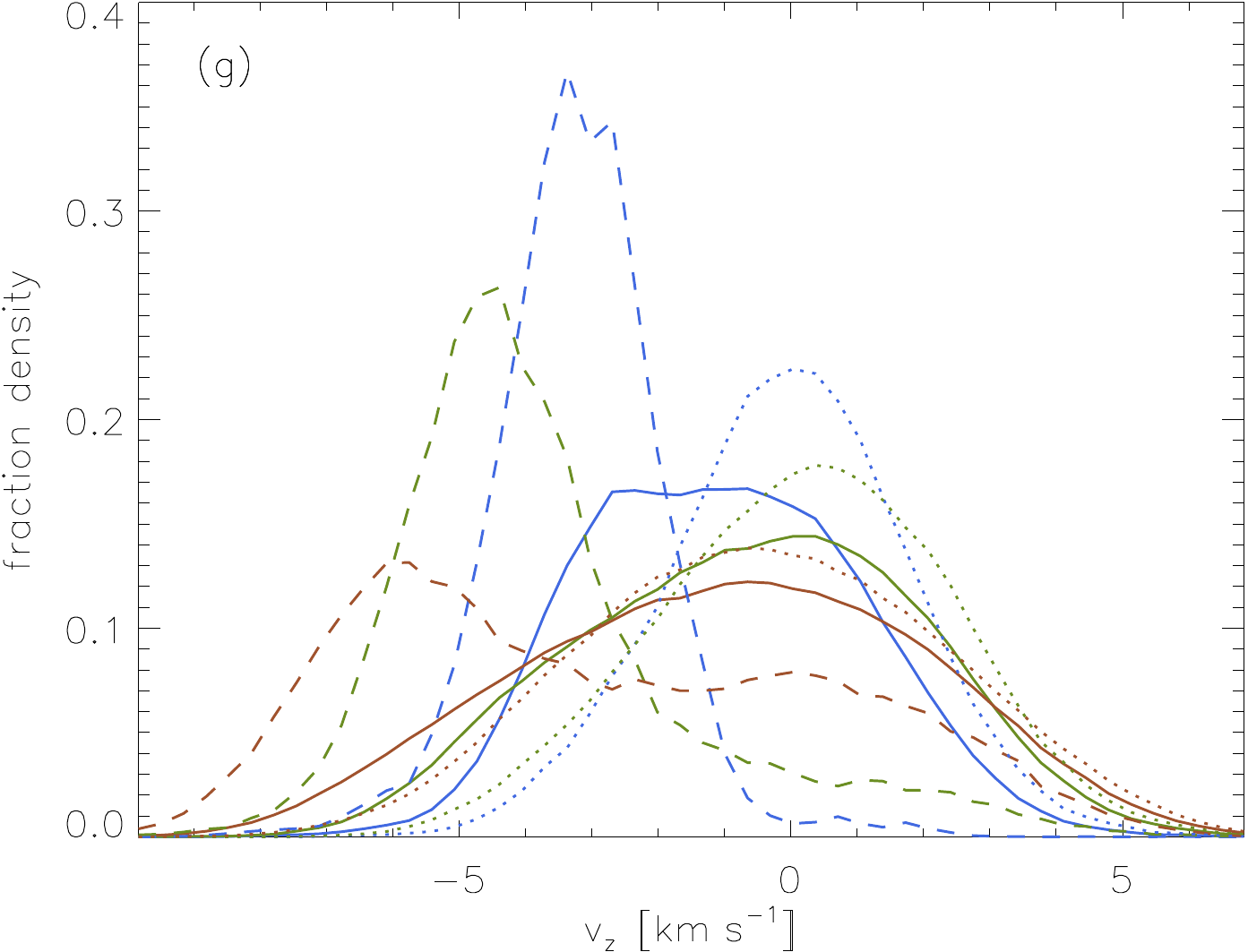} &
    \includegraphics[scale=\myscale]{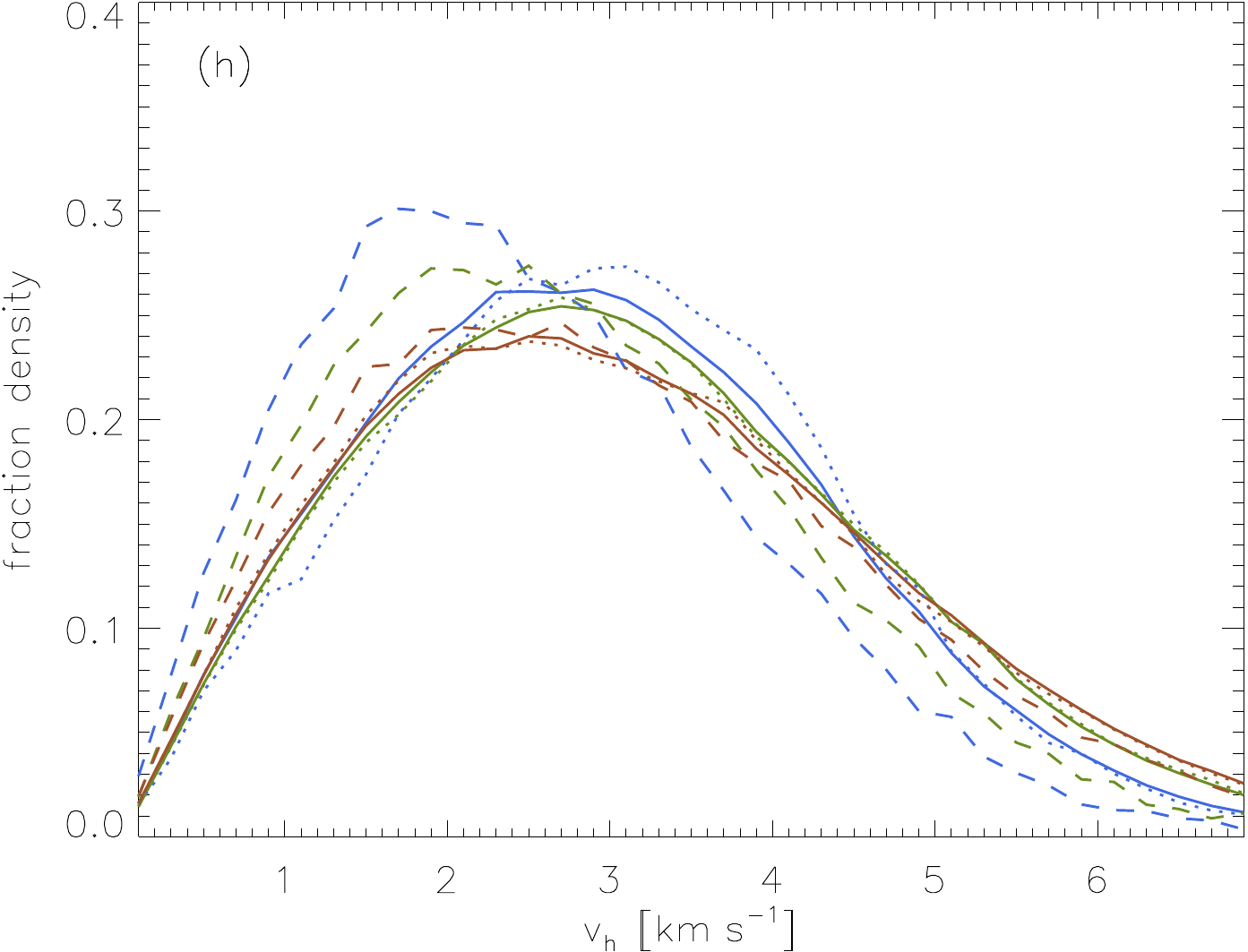}
\end{tabular}
\caption{Vortex cell statistics for Run~C.  Except for panel~(a), different
heights are represented by brown (\(z=-90\unit{km}\)), green (\(z=0\)) and blue
lines (\(z=90\unit{km}\)).  Except for panel~(b), only strong swirls with
\(\taus<120\unit{s}\) are taken into account.  The dashed and dotted lines
represent the subset of vortices with vertical (\(\iota < 20\degree\)) and
horizontal (\(\iota > 70\degree\)) orientation, respectively.  The fraction
densities plotted in panels~(b)--(f) are normalized histograms, the
normalization being such that the integral over all values of the respective
abscissa variable is one.  Angle brackets $\mean{\ldots}$ denote the horizontal
mean of the respective variable in downflow regions (i.e., where \(v_z < 0\) at
the respective height).}
\label{fig:disdens}
\end{figure*}

In all our simulations, the regions of strong swirl form a network of highly
tangled filaments. An illustrative example of this is shown in
Fig.~\ref{fig:Im3D}.  Some of the filaments protrude from the optical surface,
forming variously shaped vortex tubes with different inclinations, sometimes
bridging horizontally separated points in an arc. The optical surface is
usually depressed at the footpoints of vertically emerging vortex tubes.

The horizontal distribution of the swirls is not homogeneous, the granular
upflows being mostly devoid of them.  Near the optical surface, vortices with a
large inclination with respect to the vertical (\(\iota > 70\degree\)) are
preferentially located at the edges of the intergranular downflow lanes, see
Fig.~\ref{fig:mapvor}.  Those with a small inclination (\(\iota < 20 \degree\))
are mostly inside the lanes where the downflow is strong.

We find most of the strong vortical flows at a few hundred kilometers below the
optical surface, see Fig.~\ref{fig:depthcov}.  With an increasing upper limit
for the swirling period, the distribution declines less steeply with depth,
i.e., most of the very deep swirls are slow.  The structure at great depths is
also filamentary.  Horizontal swirls (dotted lines) are more numerous than
vertical ones (dashed lines), consistent with isotropy.  Near and above the
optical surface, the distribution of inclination angles deviates from isotropy
(see next Section).

\begin{figure}[t]
\centering
\includegraphics[width=.8\linewidth]{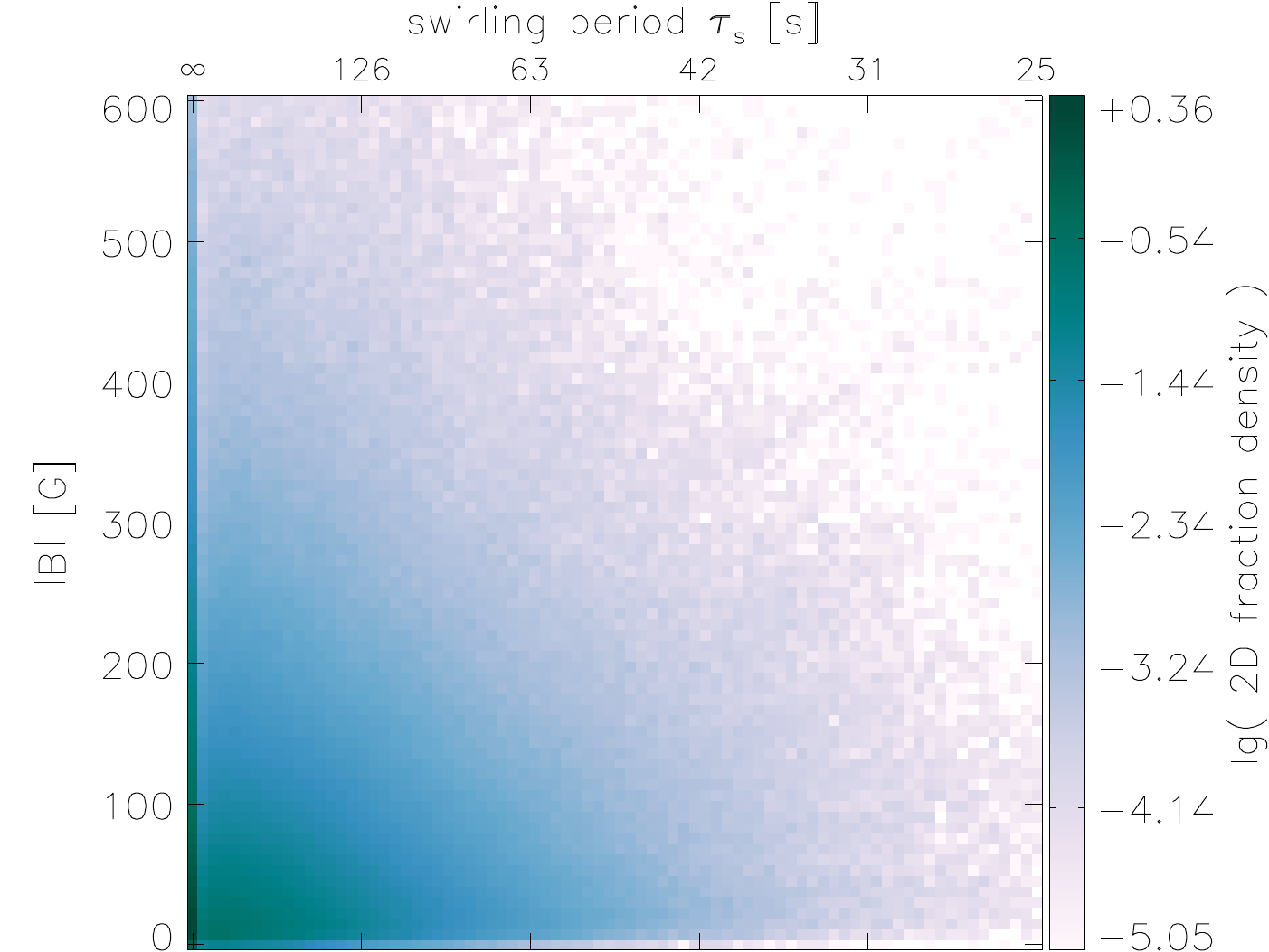} \\[\medskipamount]
\includegraphics[width=.8\linewidth]{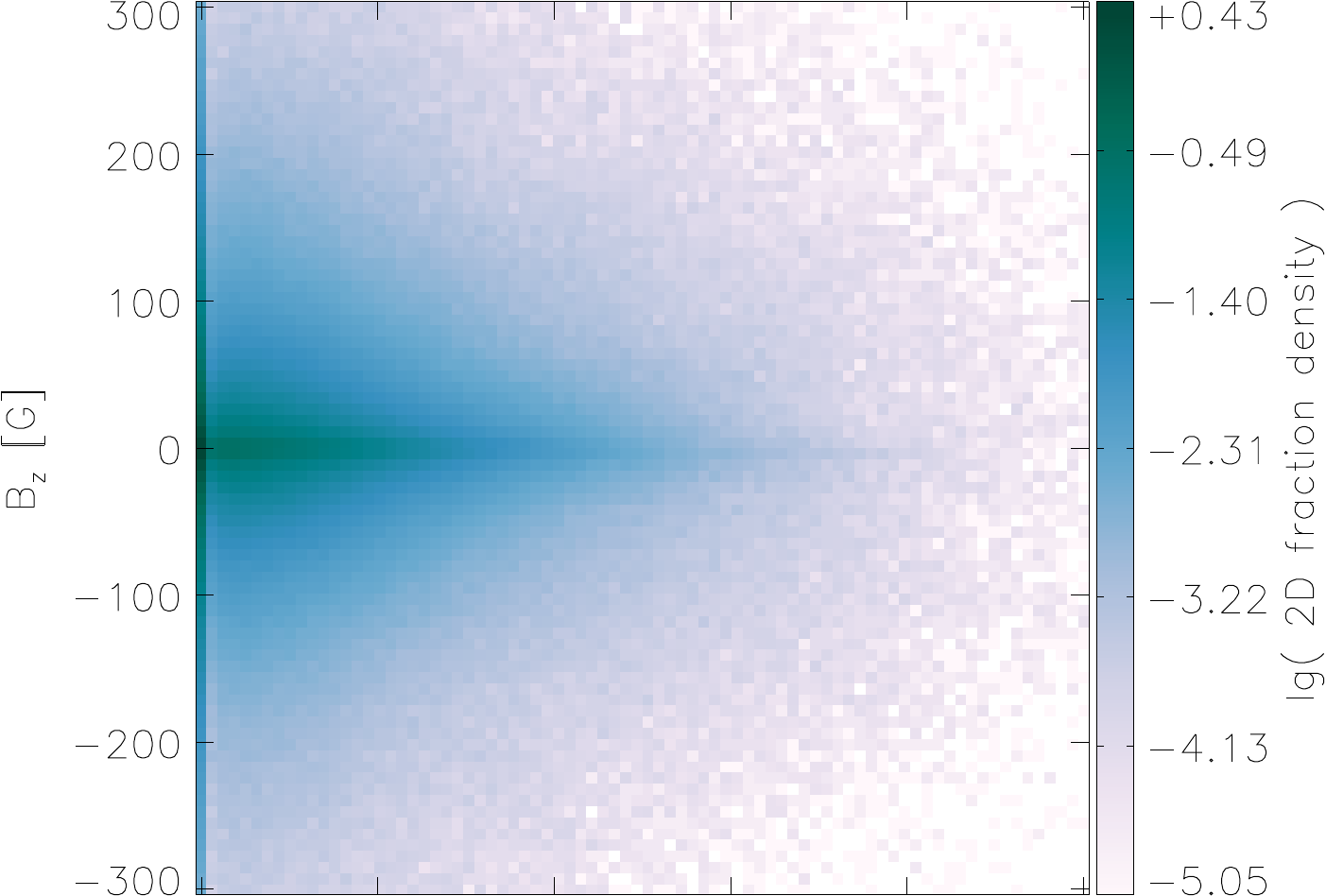} \\[\medskipamount]
\includegraphics[width=.8\linewidth]{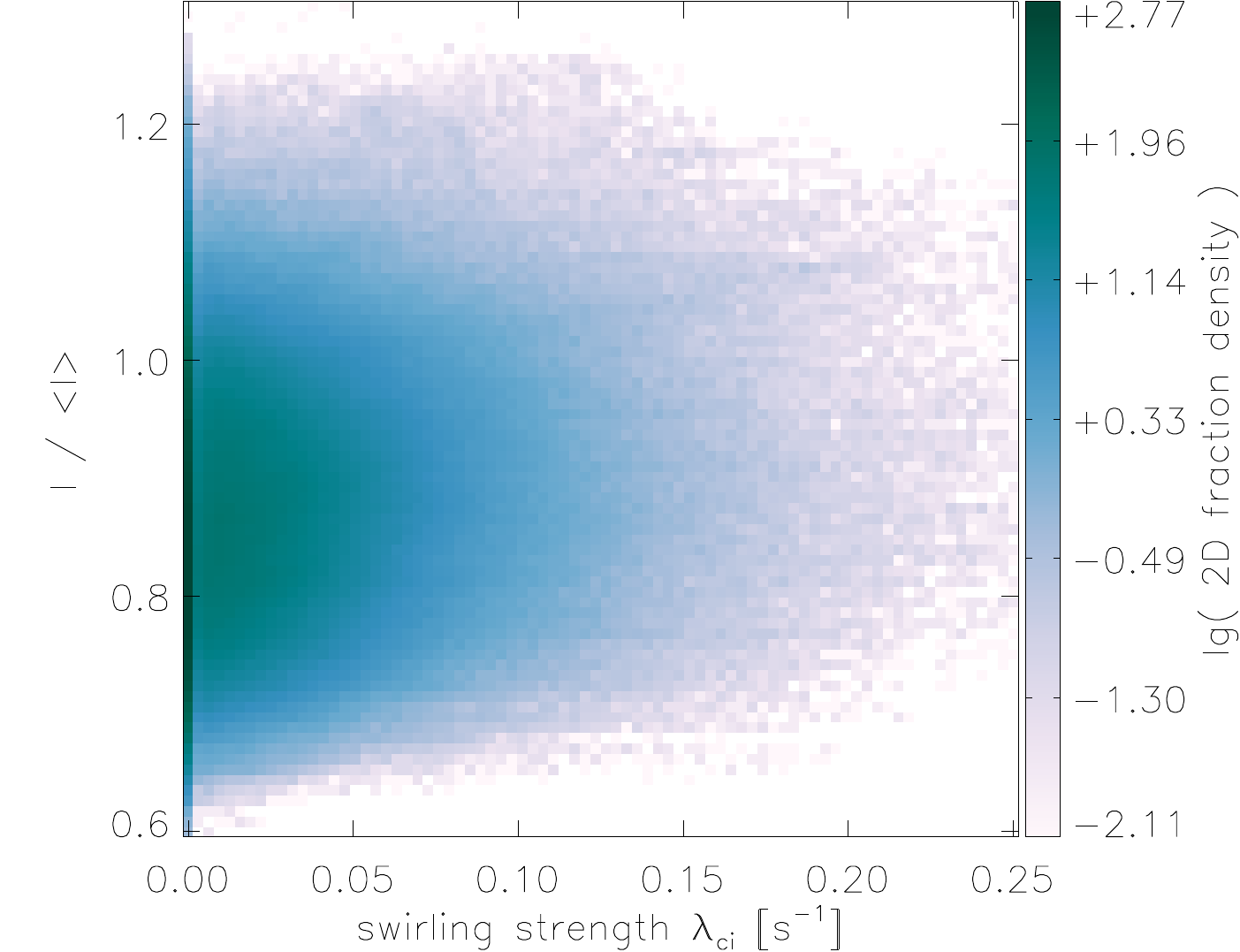}
\caption{2D histograms of the swirling strength and various observables
(\textit{top panel}: total magnetic field strength, \textit{middle panel}:
vertical magnetic field component, \textit{bottom panel}: emerging intensity
normalized by the overall horizontal mean) for Run~C. All downflow regions in a
\(\pm25\unit{km}\) range about the average height of the optical surface are
taken into account.  Each histogram is normalized such that an integral gives
the fraction of volume with properties in the integrated range.}
\label{fig:twodens}
\end{figure}

\begin{figure*}[t]
\end{figure*}

\begin{figure*}[p]
\centering
\includegraphics[width=.8\linewidth]{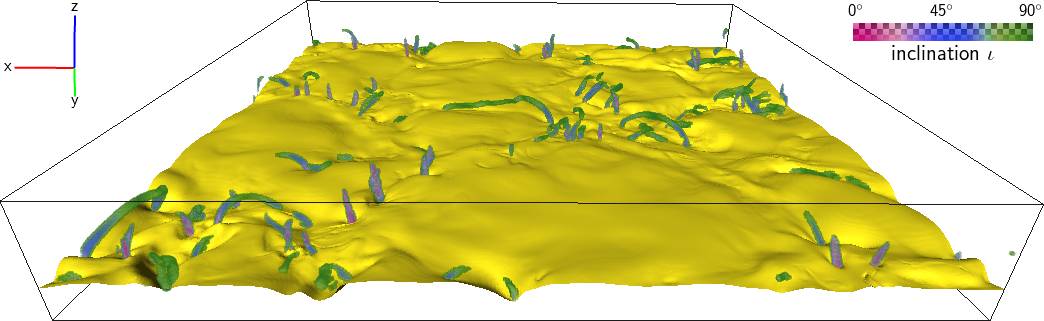}
\caption{Snapshot of Run~C with the warped optical surface in yellow and a
volume rendering of the inclination angle $\iota$ in selected vortex features
(see Sect.~\ref{sec:contiguous} for a description of the selection criteria).
Pink and green colors correspond to vertically and horizontally oriented
vortices, respectively. The size of the box shown is
\(4.8\times4.8\times0.7\unit{Mm^3}\).}
\label{fig:trova}
\centering
\begin{tabular}{l@{\hskip20pt}r}
    \includegraphics[scale=\myscale]{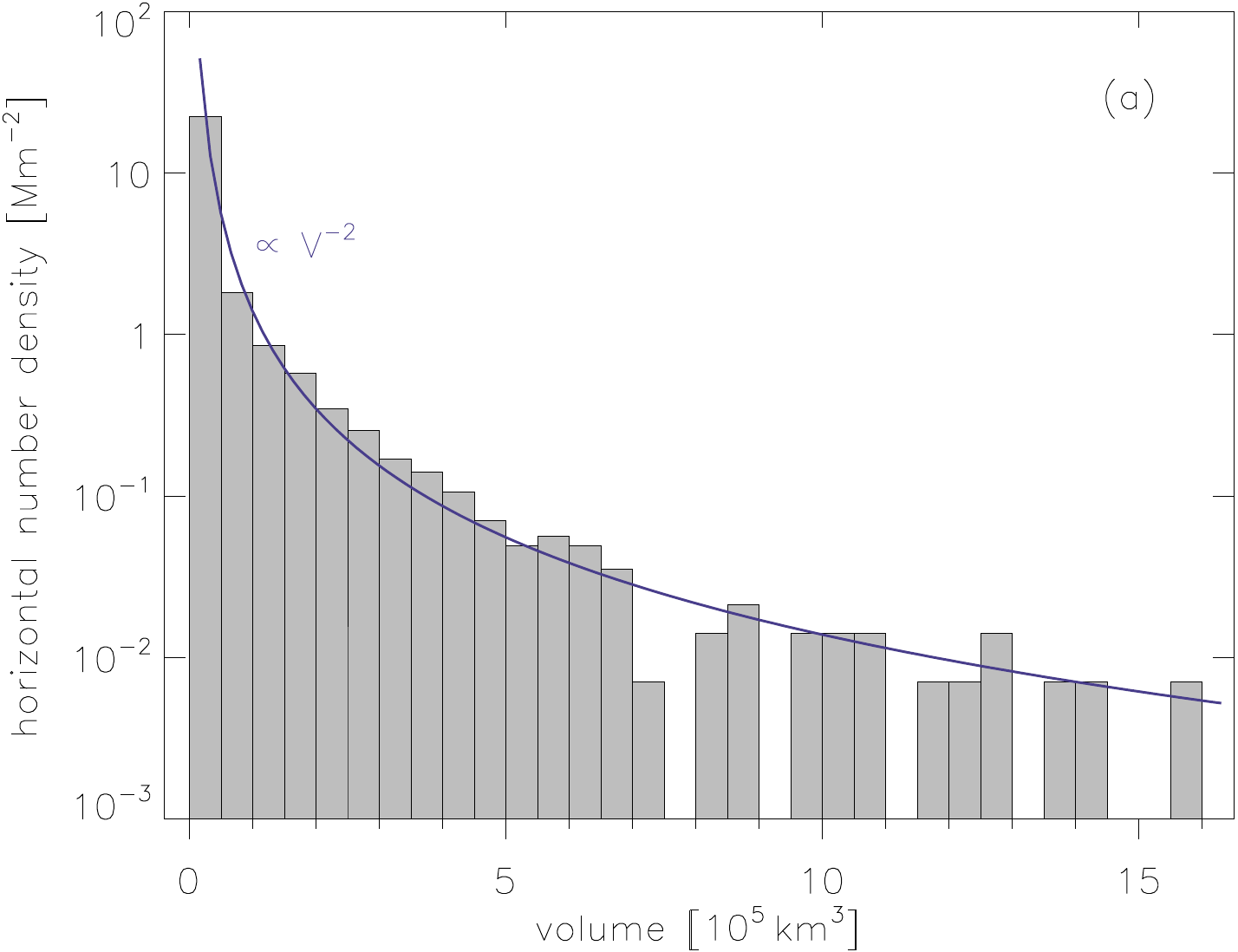} &
    \includegraphics[scale=\myscale]{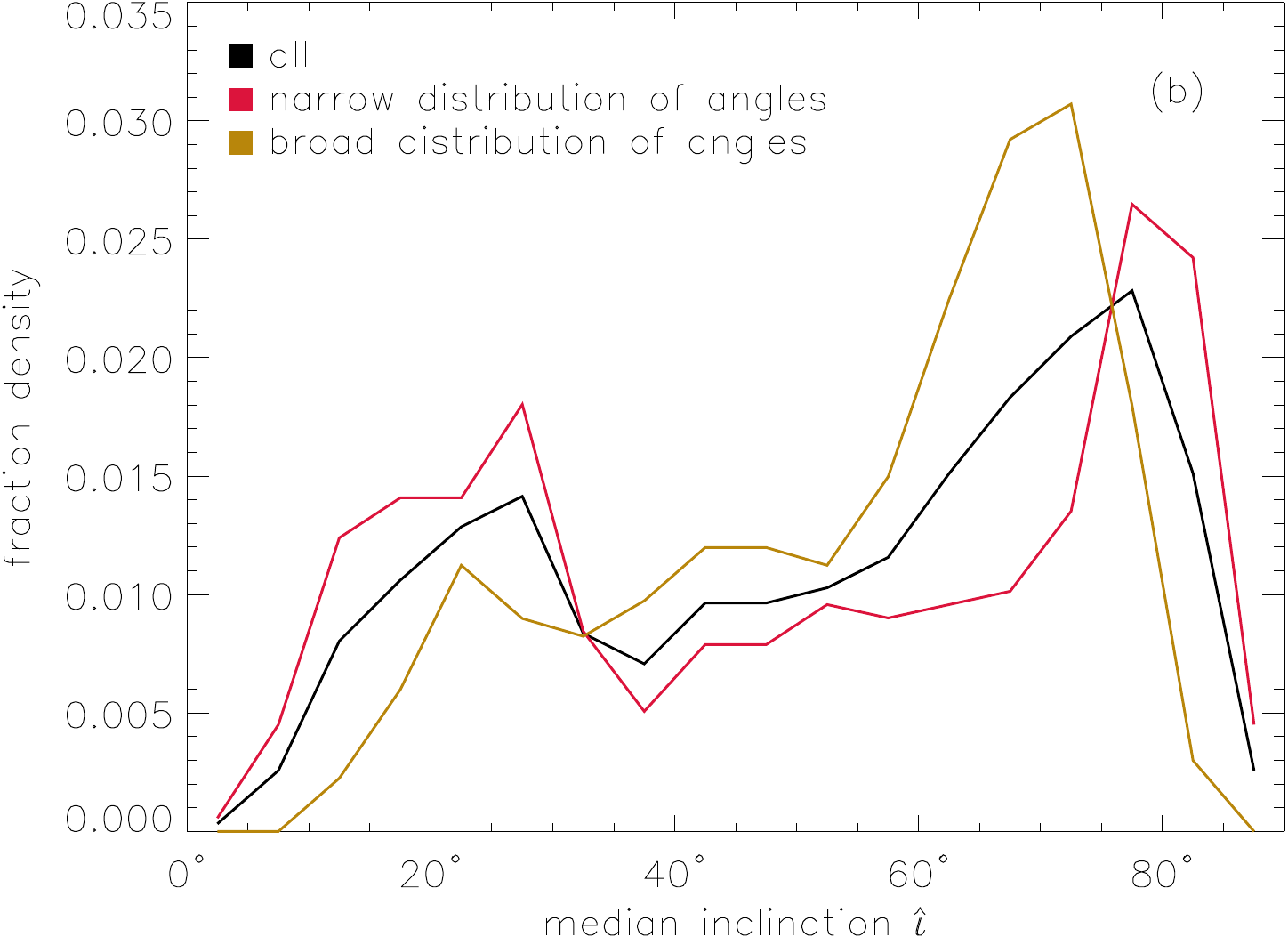} \\
    \includegraphics[scale=\myscale]{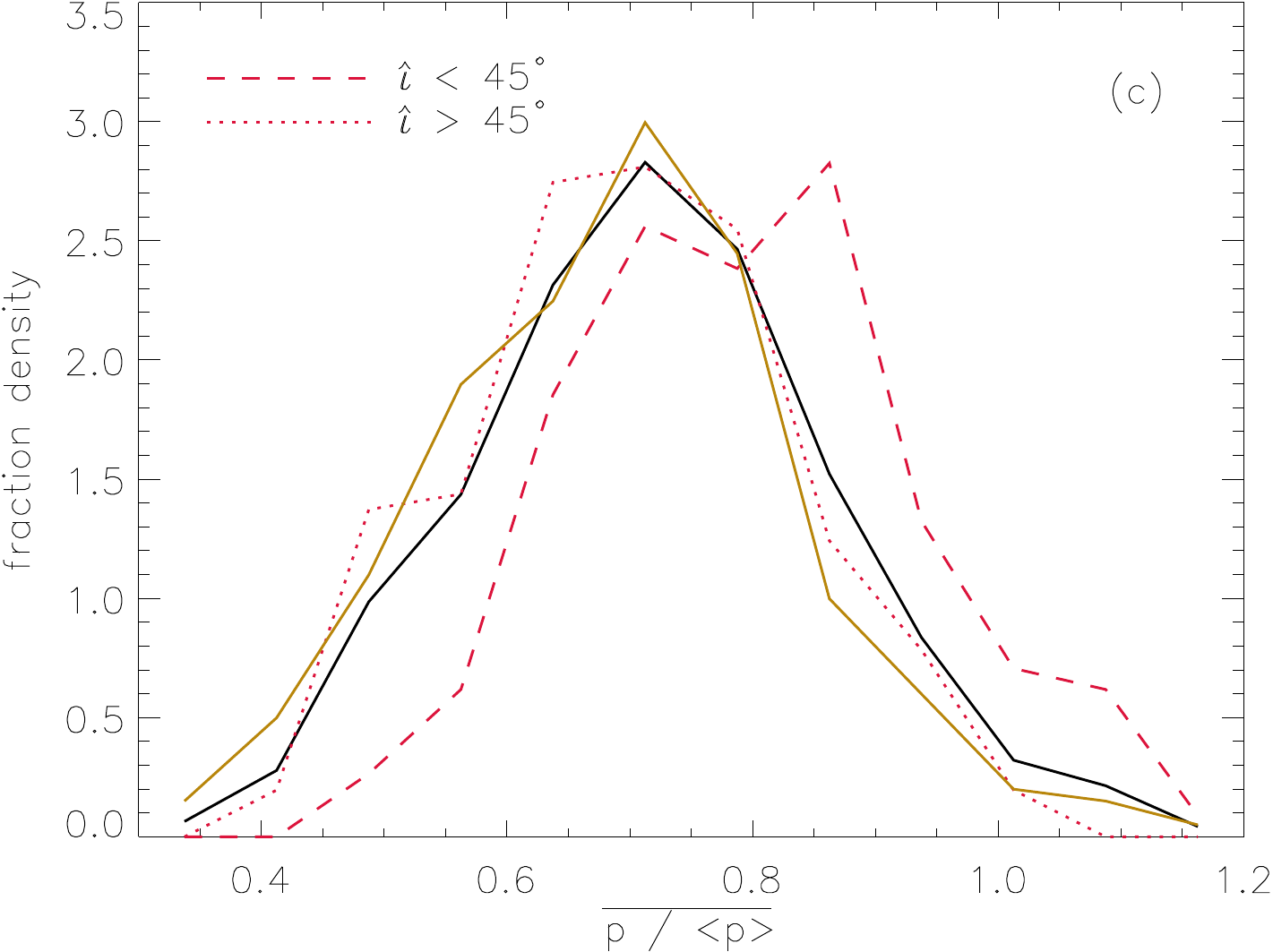} &
    \includegraphics[scale=\myscale]{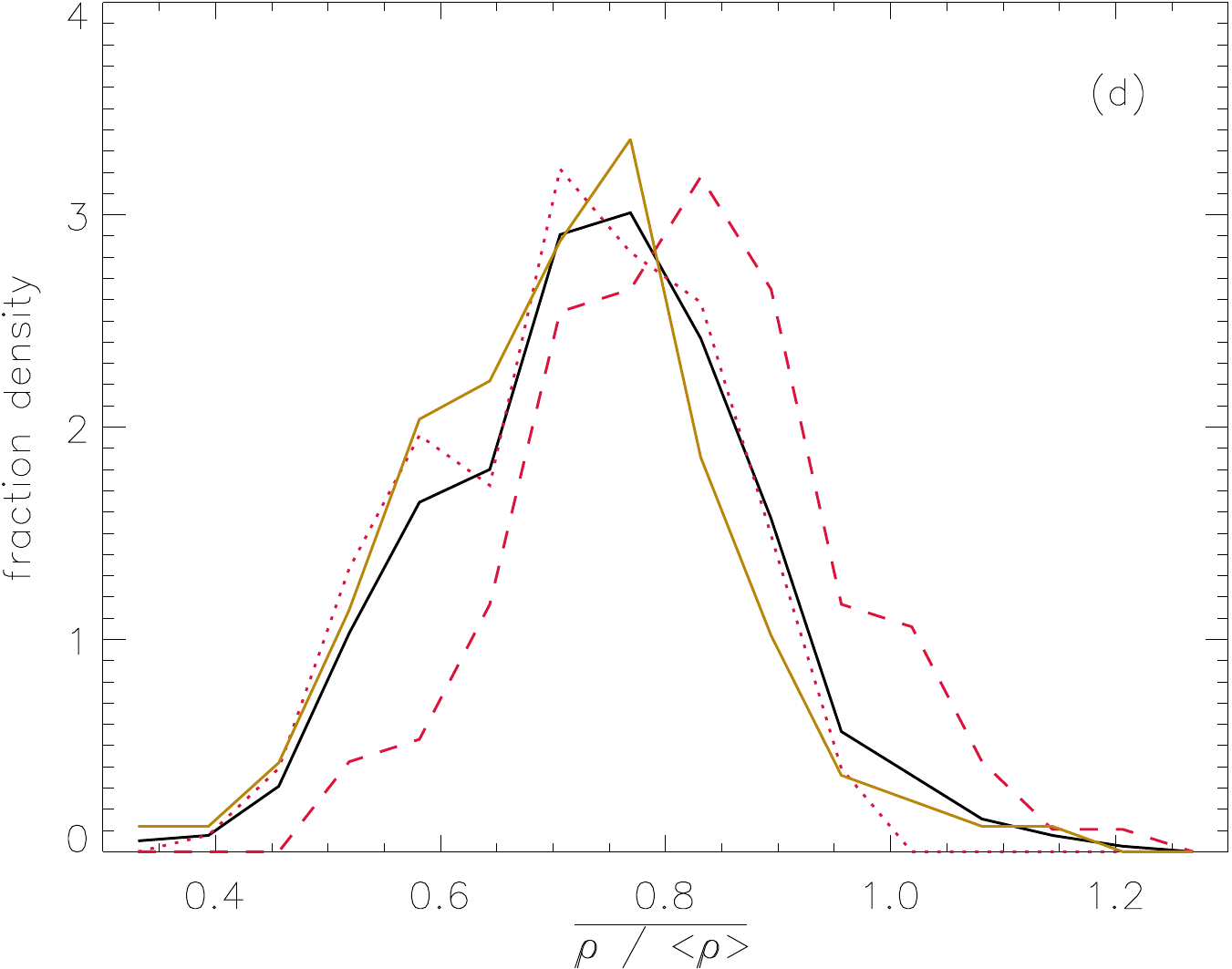} \\
    \includegraphics[scale=\myscale]{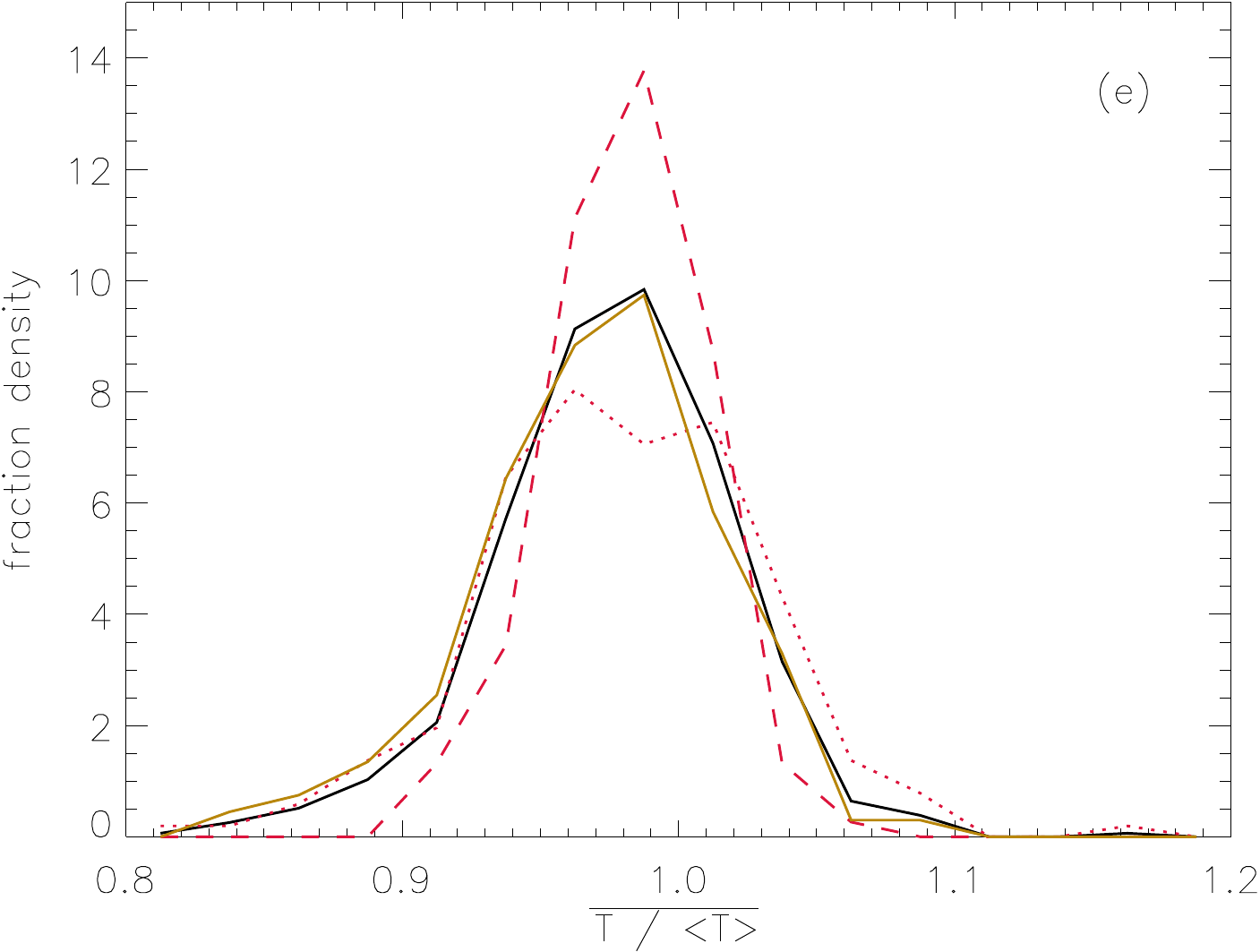} &
    \includegraphics[scale=\myscale]{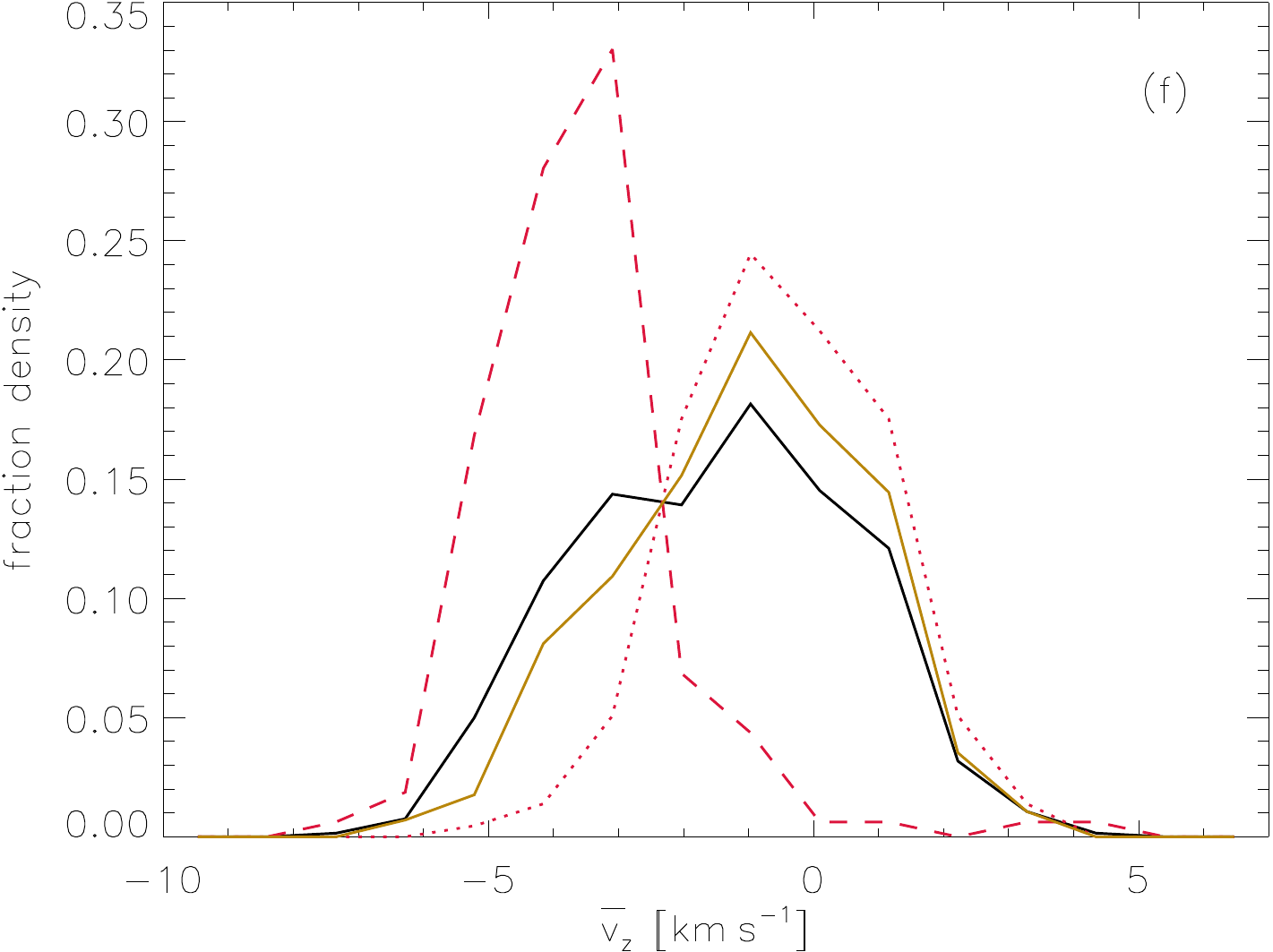}
\end{tabular}
\caption{Statistical properties of contiguous vortex features above the optical
surface in Run~C. The horizontal number density plotted in panel~(a) is the
number of features per horizontal area.  In panels~(b)--(f), features are
classified by the median inclination $\hat{\iota}$ and the width of the
distribution about the median.  Vertical (horizontal) features are deemed to be
those with narrow distributions and small (large) $\hat{\iota}$; they are
represented by dashed (dotted) lines in panels~(c)--(f). The plotted fraction
densities are histograms normalized such that the integral over all values of
the respective abscissa variable is one.  In the abscissa labels, angle
brackets \(\mean{\ldots}\) denote the (height-dependent) horizontal mean of the
respective quantity in downflow regions and a bar denotes the mean within a
particular feature.}
\label{fig:contig}
\end{figure*}

\begin{figure*}[p]
\begin{tabular}{ll@{\hskip10pt}c@{\hskip10pt}r}
\parbox[t]{.7cm}{\vspace{-3.2cm}\begin{center}time:\\[\medskipamount]$0\unit{s}$\end{center}} &
\includegraphics[scale=.288]{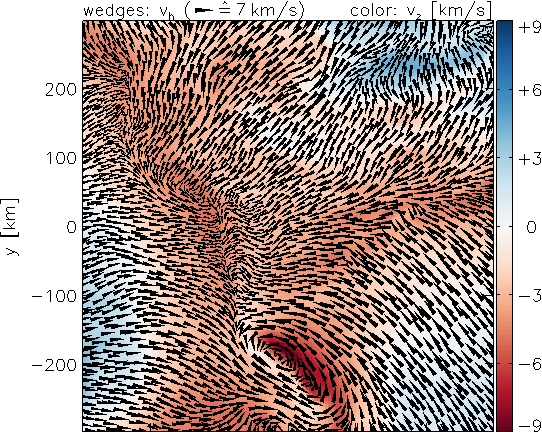} &    % scale for 72dpi
\includegraphics[scale=.4]{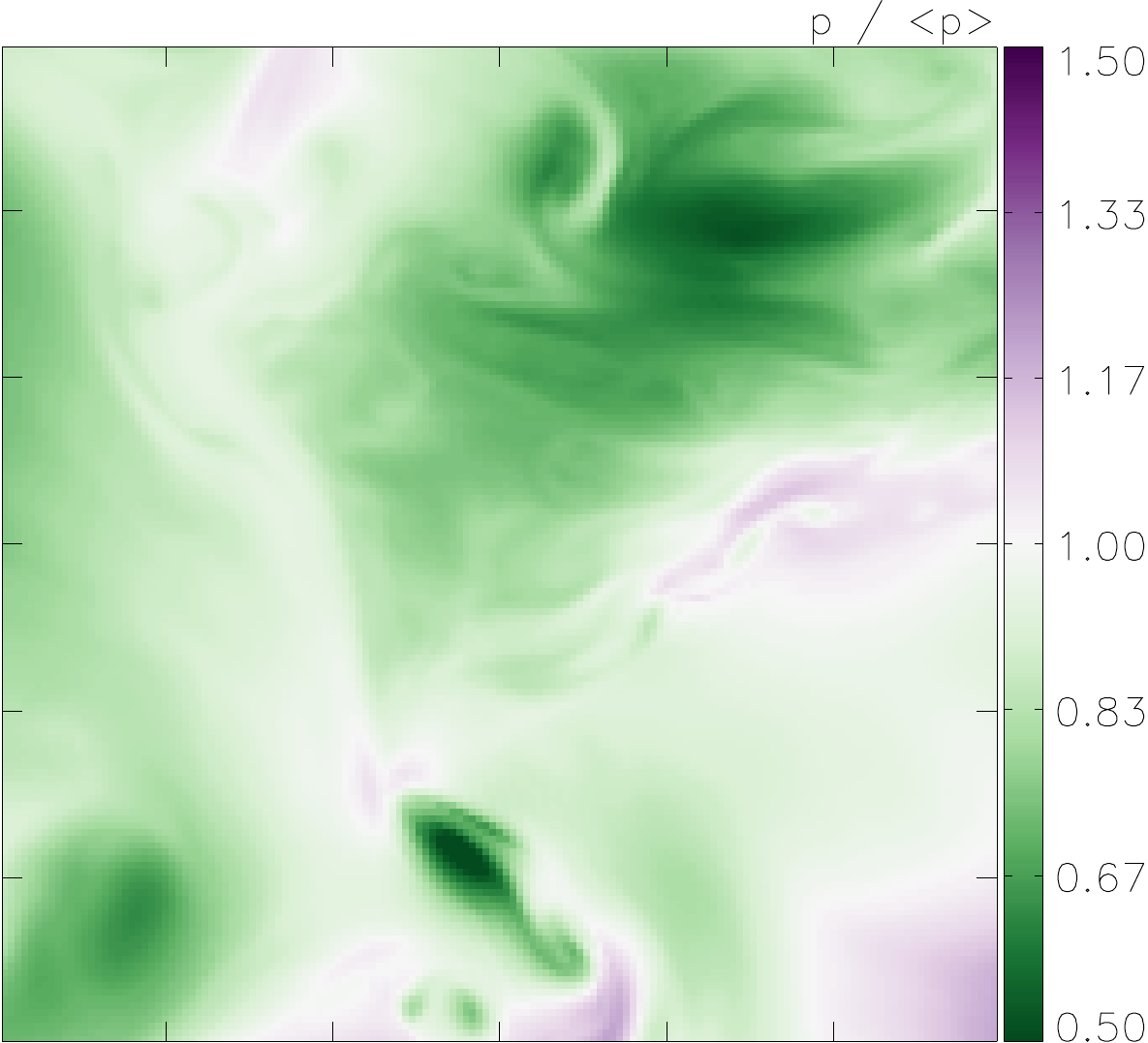} &
\includegraphics[scale=.4]{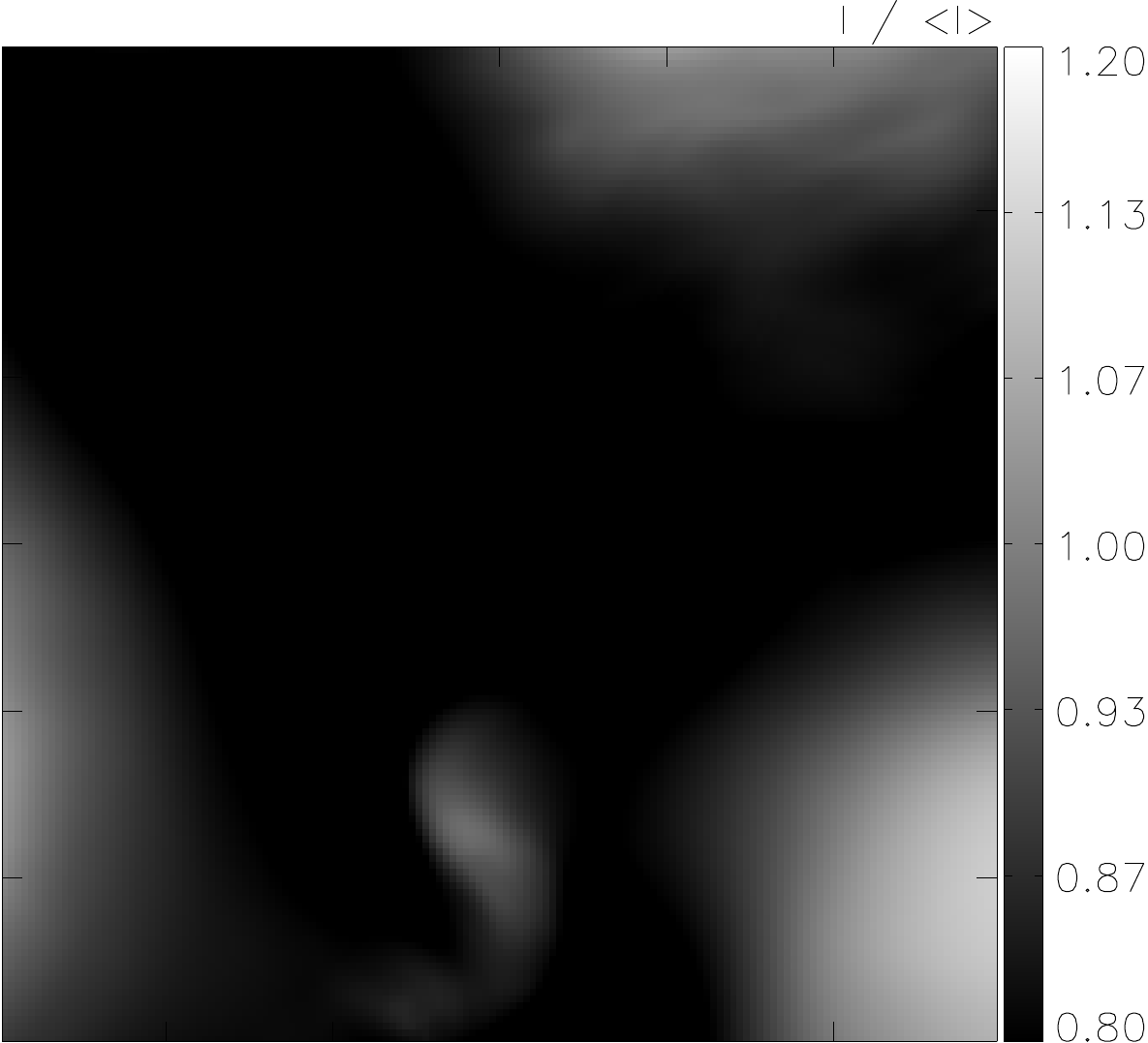} \\
\parbox[t]{.7cm}{\vspace{-2.6cm}\begin{center}$50\unit{s}$\end{center}} &
\includegraphics[scale=.288]{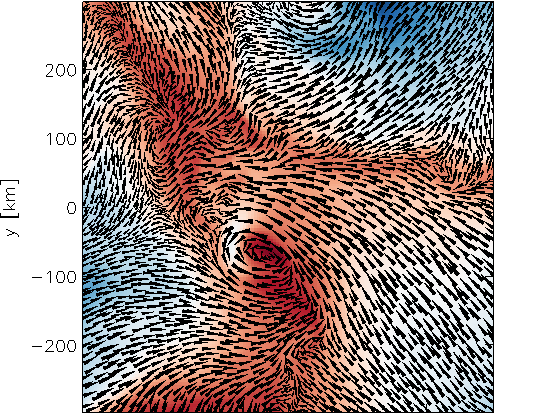} &
\includegraphics[scale=.4]{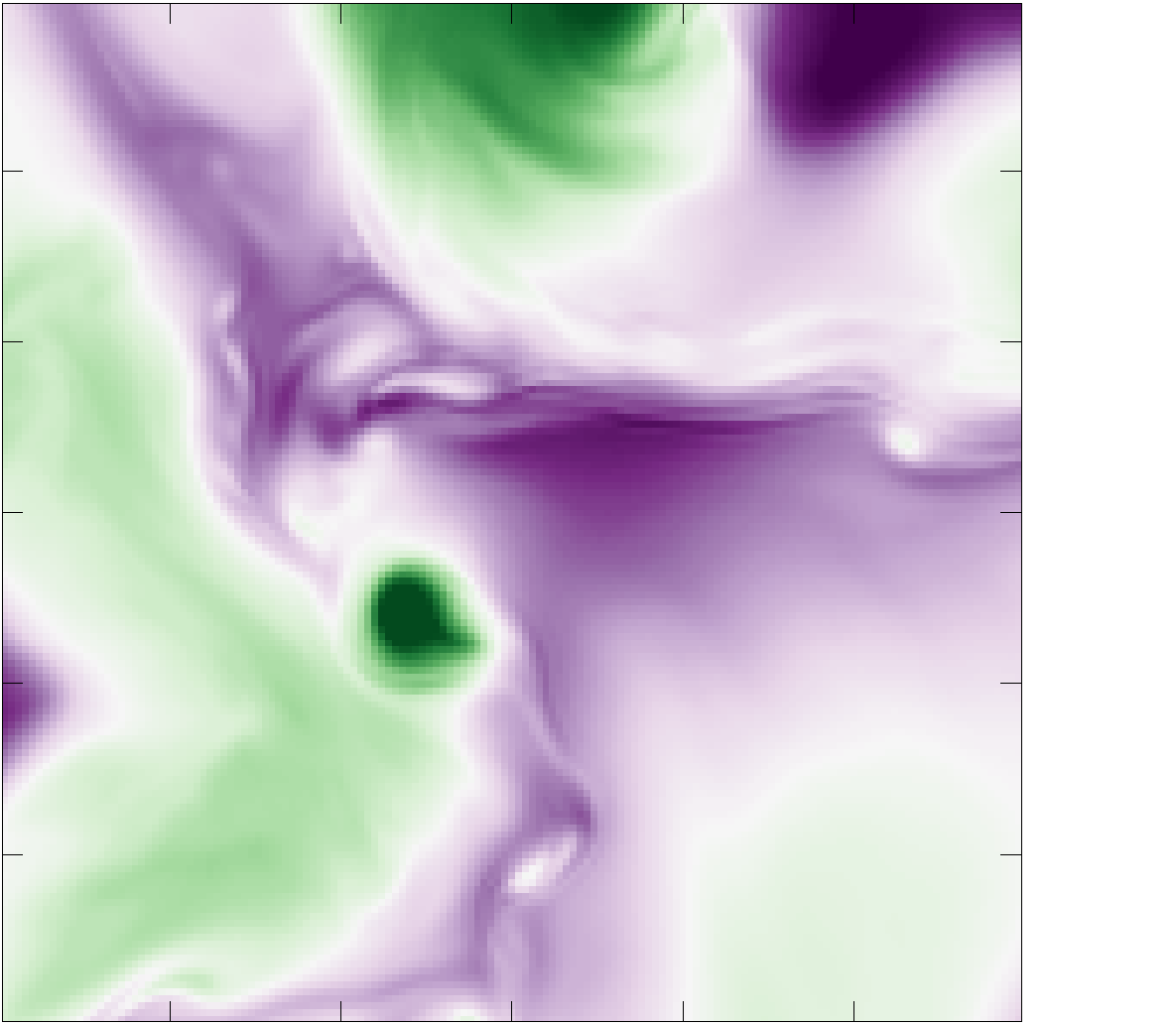} &
\includegraphics[scale=.4]{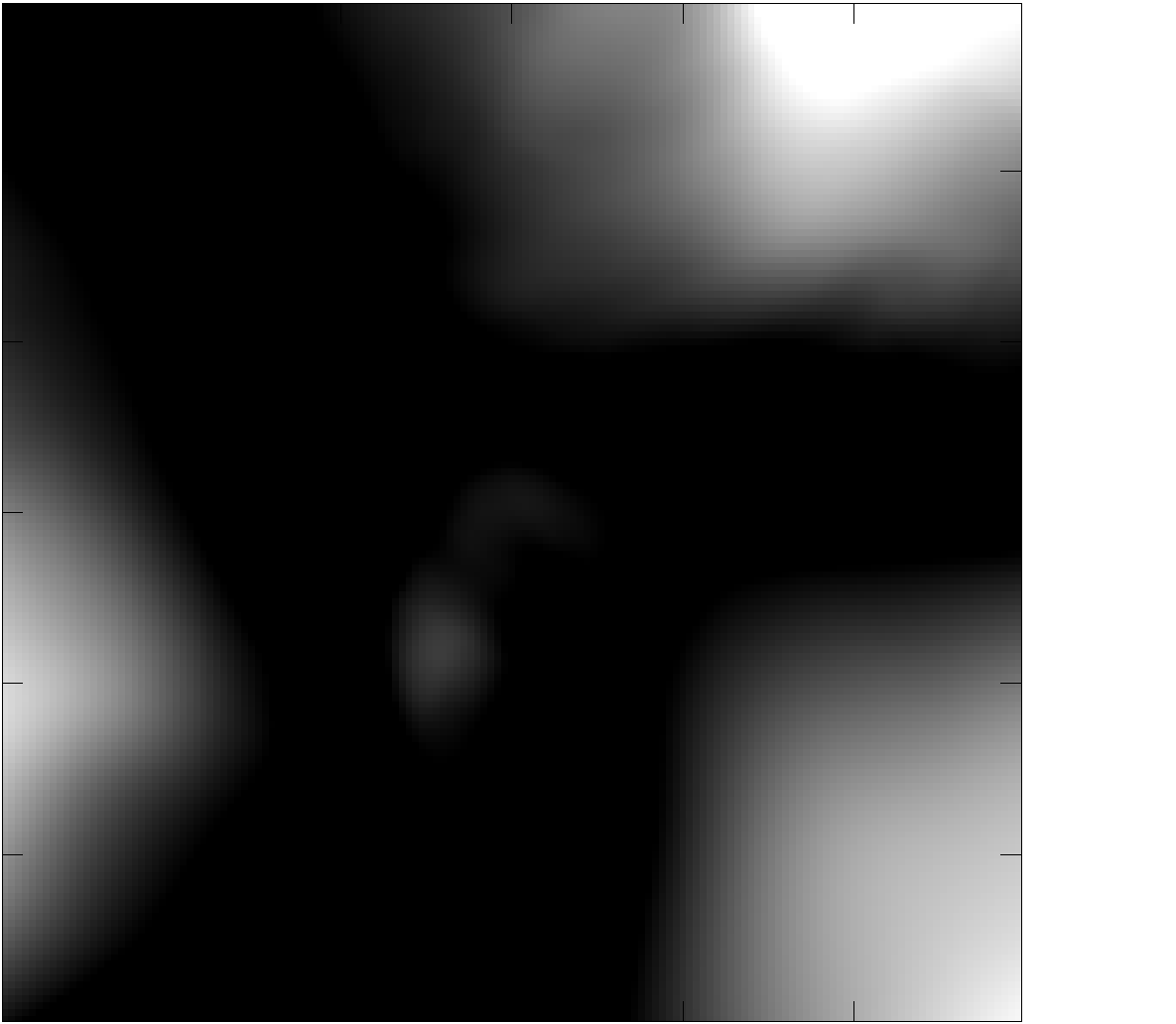} \\
\parbox[t]{.7cm}{\vspace{-3.1cm}\begin{center}$126\unit{s}$\end{center}} &
\includegraphics[scale=.288]{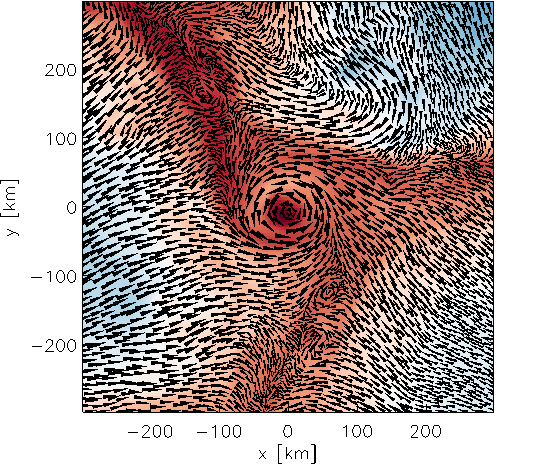} &
\includegraphics[scale=.4]{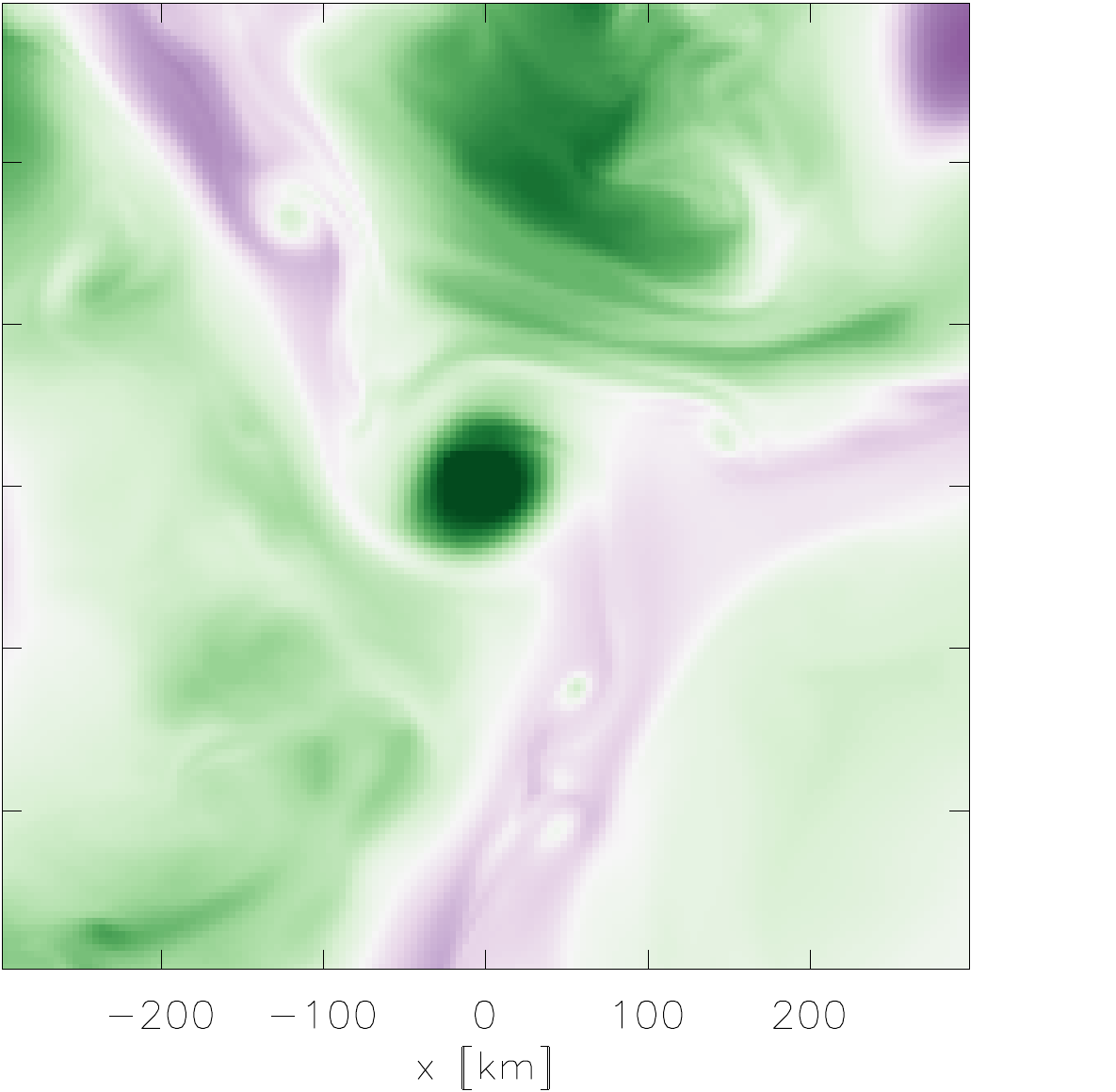} &
\includegraphics[scale=.4]{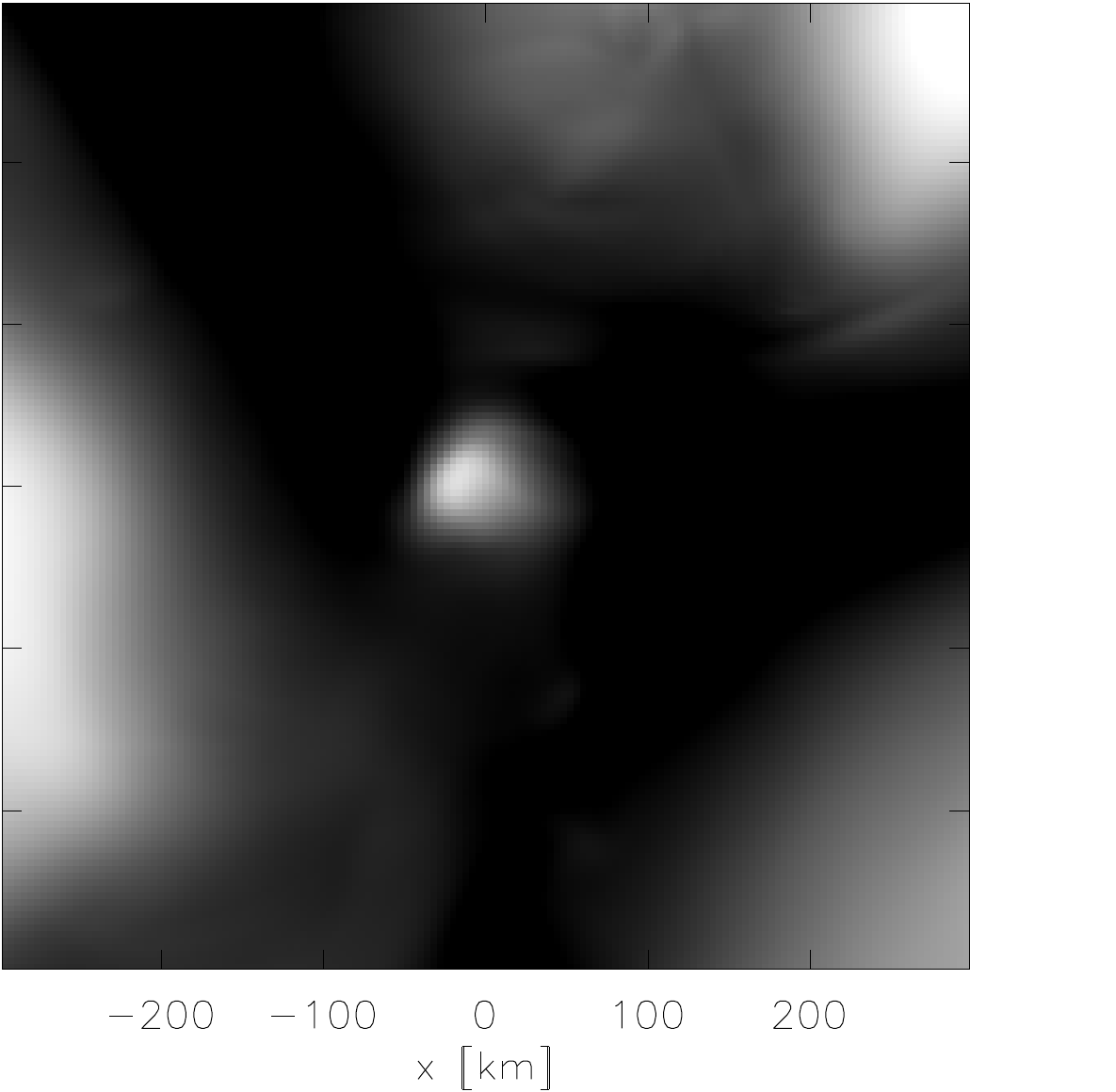}
\end{tabular} \\[-7pt]
\begin{tabular}{lr}
\imagetop{
\begin{tabular}{l}
\includegraphics[width=.3\linewidth]{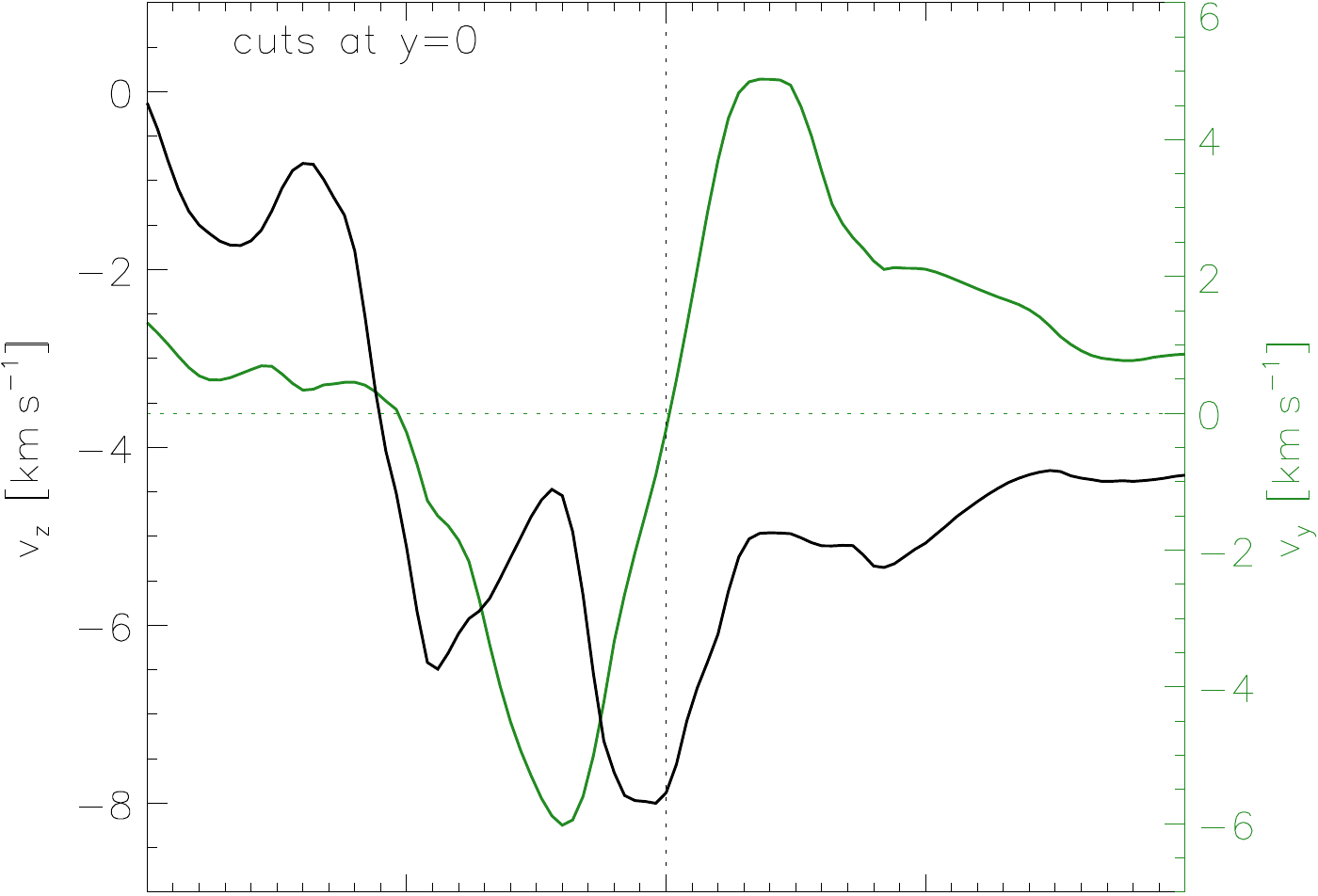} \\
\includegraphics[width=.3\linewidth]{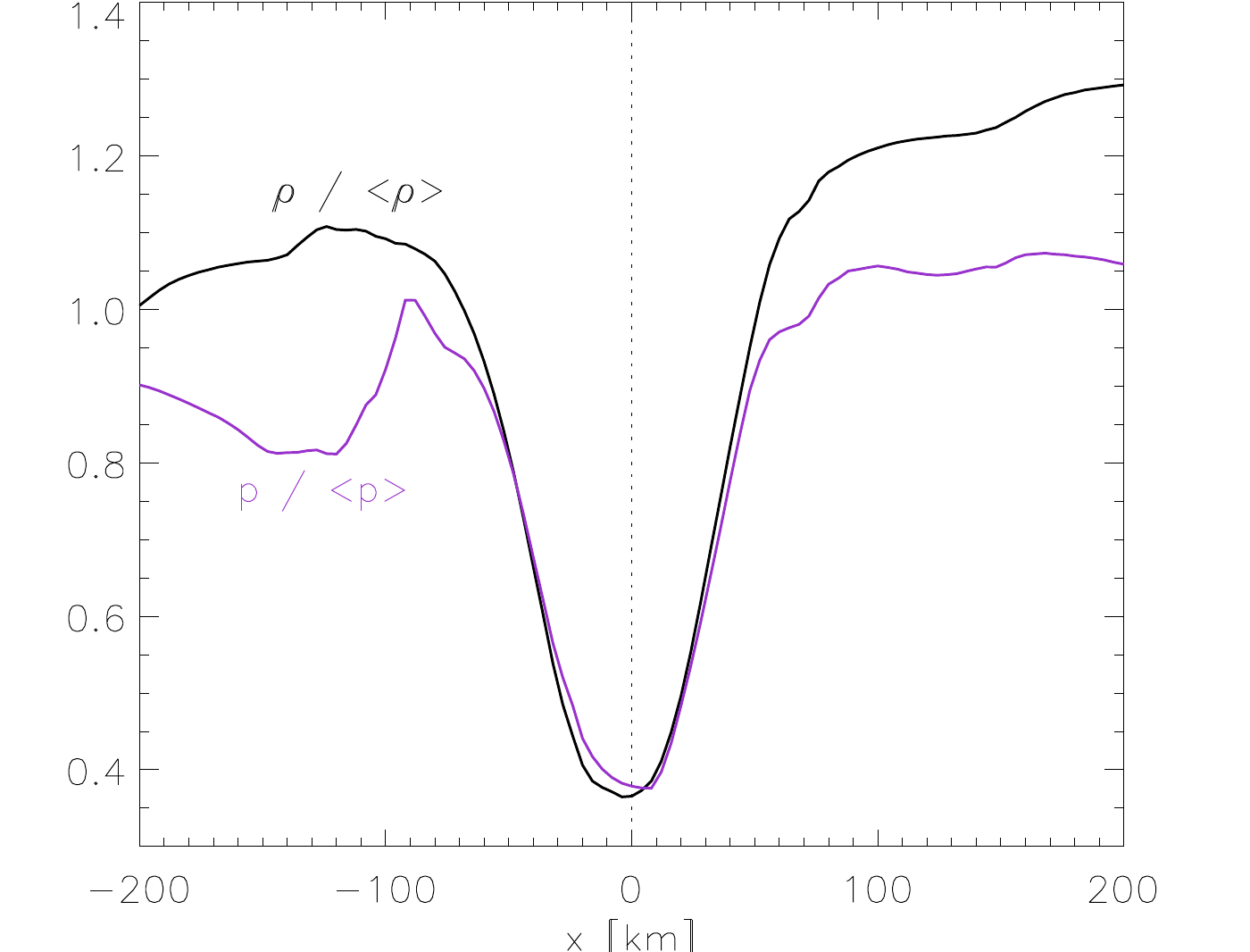}
\end{tabular}} &
\imagetop{\includegraphics[width=.59\linewidth]{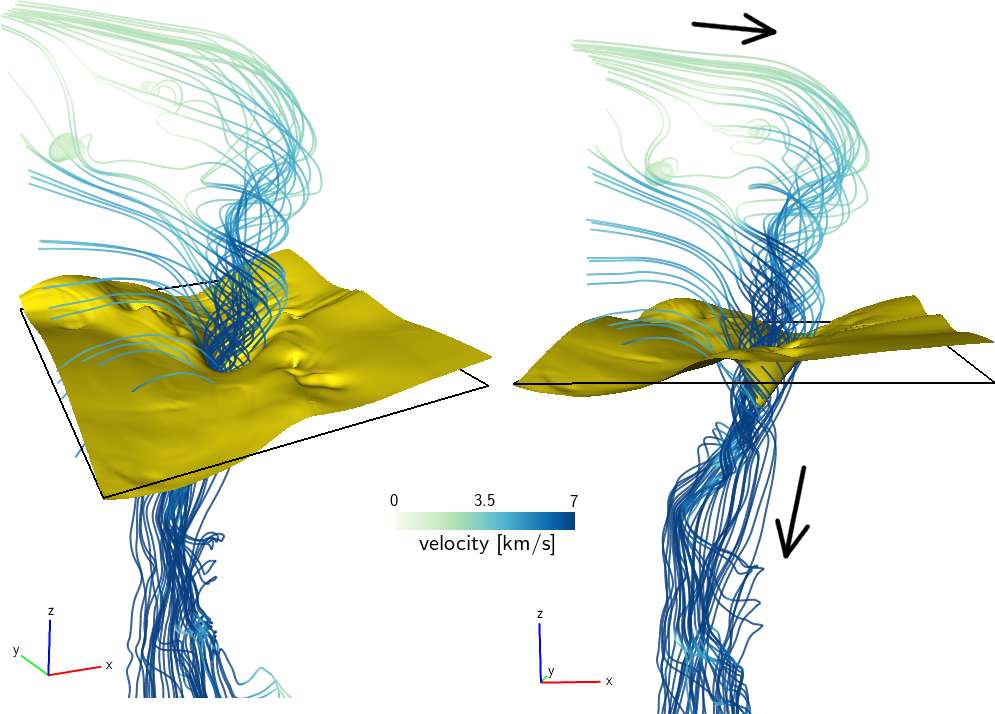}}
\end{tabular}
\caption{A strong vertical vortex penetrating the optical surface in Run~H. The
nine panels at the top show, at various stages of development, the velocity
(\textit{left panels}) and gas pressure (\textit{middle panels}) at the average
height of the optical surface (\(z=0\)) together with a bolometric intensity
map (\textit{right panels}).  The line plots at the \textit{bottom left}
represent one-dimensional cuts through the vortex at the last instant
(\(t=126\unit{s}\)). The 3D visualization at the \textit{bottom right} shows,
from two different viewing angles, selected streamlines color coded by velocity
and the warped optical surface in yellow; the black rectangle indicates the
\(z=0\) plane.}
\label{fig:manvort}
\end{figure*}

\subsection{Statistical properties of vortex regions}
\label{sec:singcellstat}

In Fig.~\ref{fig:disdens}, we present statistical properties of regions with
large $\lci$. For these analyses, we combined data from five different
snapshots of Run~C with a separation of about 7 minutes each.  All grid cells
are treated equally in the statistics, regardless of whether or not they form
part of a larger, contiguous swirling region.  In panels~(b)--(h), we plot
histograms normalized such as to give the fraction density with respect to the
considered variable: the integral over a certain range of the variable gives
the fraction of grid cells with values in that range.  For every histogram we
considered a \(50\unit{km}\) (5 grid cells) vertical range about the indicated
height, taking a total of \(\mathrm{\sim}10^7\) grid cells into account.  The
individual panels of Fig.~\ref{fig:disdens} are described in the following.

(a) This plot shows, as a function of height, the fraction of space occupied by
swirls with \(\taus<120\unit{s}\) (compare Fig.~\ref{fig:depthcov}).  The
highest occurrence of these swirls is at \(z \approx -300\unit{km}\) (\(z=0\)
being the average height of the optical surface).  Inwardly spiraling flows
that diverge in the direction of the vortex (type 2, see
Sect.~\ref{sec:vortex}) are prevailing everywhere; their mean fraction with
respect to all types is 67\% below \(z=0\) and 75\% above.  The mean fraction
of swirling flows of the outwardly spiraling, converging type (3) is 22\% below
\(z=-200\unit{km}\) and decreases to 7\% above \(z=100\unit{km}\). Types (1)
and (4) are insignificant.

(b) This plot shows histograms of the swirling period $\taus$, normalized such
that the integral over all \(\taus\) is one.  The total fraction of grid cells
with imaginary eigenvalues is about 53\% at \(z=-90\unit{km}\) and \(z=0\), and
47\% at \(z=90\unit{km}\).  With decreasing height, the histogram is more
peaked and the peak moves toward larger swirling strengths (smaller $\taus$).
In all the other plots of Fig.~\ref{fig:disdens}, only strong swirls with
\(\taus < 120\unit{s}\) (i.e., to the left of the vertical line) are taken into
account.

(c) The histograms of the inclination angle $\iota$ indicate that the swirls
are preferentially (i.e., more than expected for an isotropic distribution)
horizontal at \(z=0\) and preferentially horizontal or vertical at
\(z=90\unit{km}\), whereas at \(z=-90\unit{km}\) the distribution is consistent
with an isotropic distribution (indicated by the gray dash-dotted line).

(d) The gas pressure $p$ in swirling regions is, on average, reduced to 87\% of
the horizontal mean \(\mean{p}\) in all downflow regions (\(v_z<0\)) at the
same height, consistent with a contribution by a dynamic pressure due to
centrifugal forces. The gas pressure deficit does not depend much on height and
inclination angle.

(e) The swirling regions are significantly rarefied, the average value of
\(\rho/\mean{\rho}\) being 84\%.  The rarefaction does not depend much on
height and inclination angle.

(f) The mean of the relative temperature \(T/\mean{T}\) is \(1.0\) at all three
considered heights. With increasing depth, the distribution becomes broader and
the maximum moves towards lower temperatures.  Most of the vertically oriented
swirls (dashed lines) are slightly cooler than the horizontal ones (dotted
lines), consistent with their predominant location inside cool intergranular
lanes.

(g) In horizontal swirls (dotted lines), the median and the mean of the
vertical velocity are near zero, in agreement with them being found at the
edges of the granules.  In vertical swirls (dashed lines), the mean is roughly
\(-3\unit{km\,s^{-1}}\) at all heights.

(h) The mean of the horizontal velocity \(v_\mathrm{h}\) is approximately
\(3\unit{km\,s^{-1}}\), irrespective of the height.  Note that this velocity
includes the horizontal proper motion of the whole swirl.  Vertical swirls tend
to have slightly smaller horizontal velocities, presumably because most of them
(being at the center of intergranular lanes) are not subject to the bulk motion
of the expanding granules.

In Fig.~\ref{fig:twodens}, we present 2D histograms of the swirling strength
and selected observables, taking into account all downflow regions near the
optical surface. This restricts the analysis to the intergranular lanes.  The
top panel shows that strong magnetic fields become increasingly rare with
increasing swirling strength, there being no apparent correlation between the
two. This is also true for both polarities of the vertical magnetic field
alone.

As is to be expected for the downflow lanes, the emerging intensity in vortex
regions is preferentially smaller than the overall mean (bottom panel of
Fig.~\ref{fig:twodens}).  There is no clear trend towards higher intensity with
increasing swirling strength.  This indicates that, although vortices are
locally bright, the contrast is small compared to the brightness variations
within intergranular lanes.

\begin{figure*}[t]
\includegraphics[width=\linewidth]{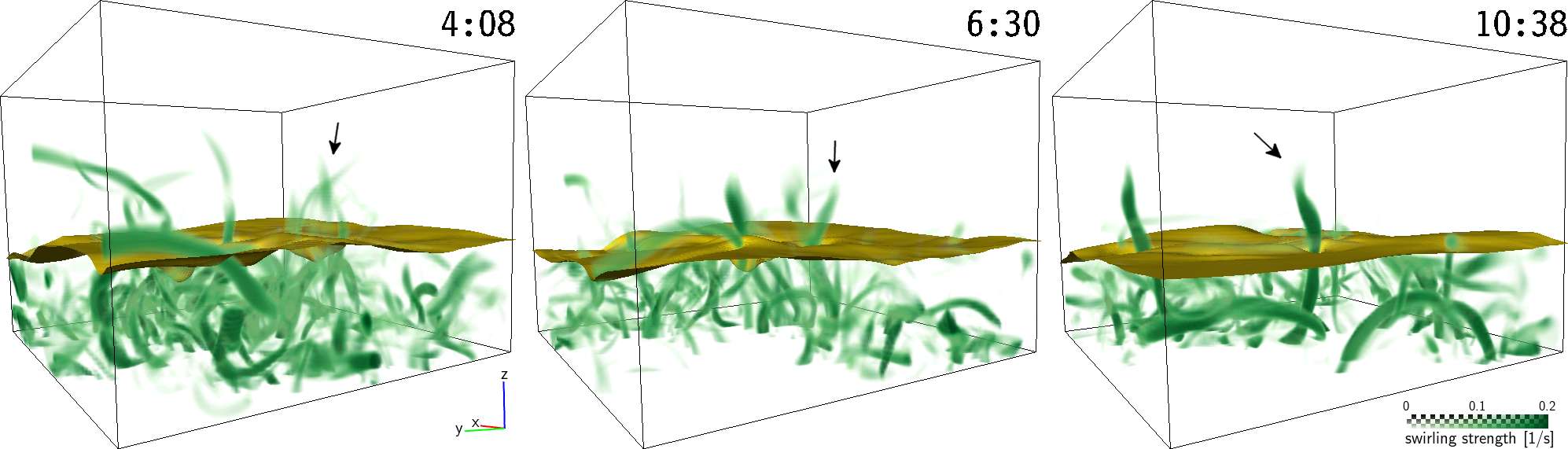}
\caption{Formation of a strong vertical vortex in Run~C. The plots display the
swirling strength (green volume rendering) and the optical surface (yellow) at
three different times (labels are in minutes). The size of the box shown is
\(1.1\times1.1\times0.8\unit{Mm^3}\).}
\label{fig:favevol}
\end{figure*}

\begin{figure*}[t]
\includegraphics[width=\linewidth]{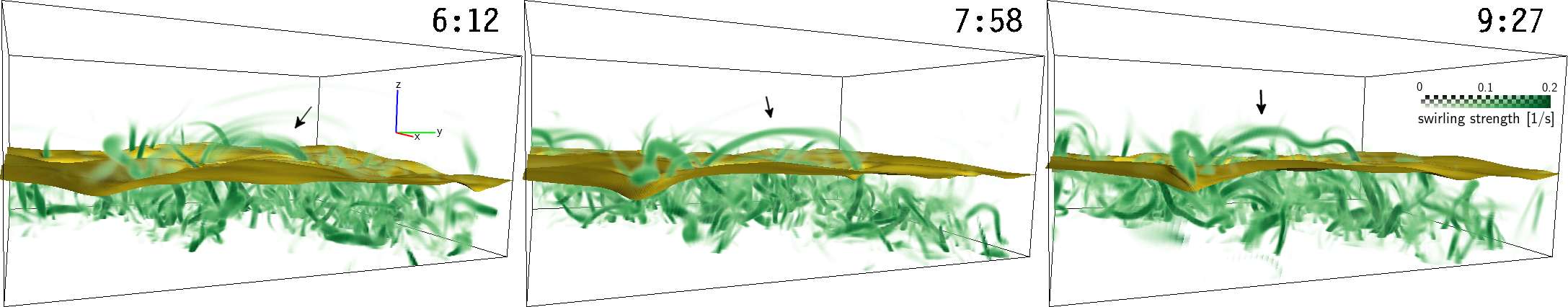}
\caption{Rise and fall of a vortex arc in Run~C. The plots display the swirling
strength (green volume rendering) and the optical surface (yellow) at three
different times (labels are in minutes). The size of the box shown is
\(1.5\times1.5\times0.8\unit{Mm^3}\).}
\label{fig:bogevol}
\end{figure*}

\subsection{Contiguous vortex features}
\label{sec:contiguous}

We define vortex features to be contiguous regions of grid cells with a large
swirling strength.  To achieve a clear separation,  we consider only features
above the (warped) optical surface and take a lower cutoff for the swirling
period than that used in the preceding section: \(z > z(\tau=1)\) and \(\taus <
60\unit{s}\).  In 6 independent snapshots of Run~C we find 3806 features that
satisfy these criteria (this number includes isolated grid cells).  The number
of features decreases rapidly with size, see panel (a) of
Fig.~\ref{fig:contig}.  It is roughly proportional to the inverse square of the
occupied volume (blue line).  In the following, we consider only the 622
features with at least 100 grid cells, corresponding to a minimum volume of
\(5.6\cdot10^4\unit{km^3}\).

Fig.~\ref{fig:trova} shows the features detected in one of the snapshots.  On
visual inspection, we find that the inclination angle $\iota$ is in alignment
with the longitudinal directions of the features, i.e., vertical (horizontal)
features consist of swirls with small (large) inclination angles and more
complicated shapes contain swirls with different inclinations.  We use the
median of the inclination angle $\iota$ as a characteristic value for each
feature. This median is denoted with $\hat{\iota}$ in the following.

The histogram of $\hat{\iota}$ for all features is represented by the black
line in panel~(b) of Fig.~\ref{fig:contig}.  The red line represents the subset
of features with a narrow distribution of angles, defined to be those for which
70\% of all the constituent cells have an inclination $\iota$ within
\(\mathord{\pm}10\degree\) about the feature's median $\hat{\iota}$.  The
golden line represents all other features.

We divide the narrow subset (red) into two groups, a vertical one for which
$\hat{\iota}$ is smaller than $45\degree$ (dashed lines) and a horizontal one
for which $\hat{\iota}$ is larger than $45\degree$ (dotted lines).  Density
functions of the gas pressure, density, temperature and vertical velocity are
plotted in panels~(c)--(f); each quantity is divided by the horizontal mean in
downflow regions at the respective height and then the mean within the
respective feature is considered.  The results are consistent with the
``point-by-point'' statistics presented in Sect.~\ref{sec:singcellstat}
(compare with Fig.~\ref{fig:disdens}): gas pressure and density are
significantly reduced with respect to the horizontal mean, the differences
between the horizontal and vertical subset being qualitatively similar.  The
vertical velocity is strongly negative (downward) in almost all of the vertical
features.

\paragraph{Lifetimes}
Through visual inspection, we estimate the lifetimes of 17 features in Run~C.
We define the lifetime to be the time interval in which a feature is visible as
a distinct entity above the optical surface. The resulting mean lifetime is
3:30 minutes, the standard deviation being 1:40 minutes.

\subsection{Individual examples and evolution}

\subsubsection{Vertical vortices}

Fig.~\ref{fig:manvort} displays an example of a vertically oriented vortex
which penetrates the optical surface.  Starting out as a somewhat distorted
vortical flow (in the box's frame of reference), it evolves into a roundish
vortex with a diameter of approximately \(80\unit{km}\).  The rotation in the
horizontal plane is nearly rigid at its core (innermost \(60\unit{km}\)) and
the vertical velocity reaches a local maximum at the center of the vortex.
Both gas pressure and density are lowered to as little as 36\% of the
respective horizontal mean at \(z=0\) and are significantly distinct from the
local background.  The optical surface is depressed inside the vortex, the
bottom of the depression being \(110\unit{km}\) below \(z=0\).  At the site of
the vortex, the bolometric intensity is locally increased by up to 24\% with
respect to the mean. The vortex thus appears as a bright spot within the dark
intergranular lane.

Fig.~\ref{fig:favevol} illustrates the formation of a vertical vortex.  The
feature is indicated by a black arrow in the three snapshots shown.  It starts
as a conglomeration of weakly swirling structures that apparently merge into a
single distinct feature.  The swirling strength increases with time. At all
stages, the optical surface is depressed in the vortex, the largest depression
being on the order of \(\simm100\unit{km}\).  Above the optical surface, the
feature has a length of \(\simm350\unit{km}\).

\subsubsection{Vortex arcs}

\begin{figure}[t]
\centering
\includegraphics[width=.9\linewidth]{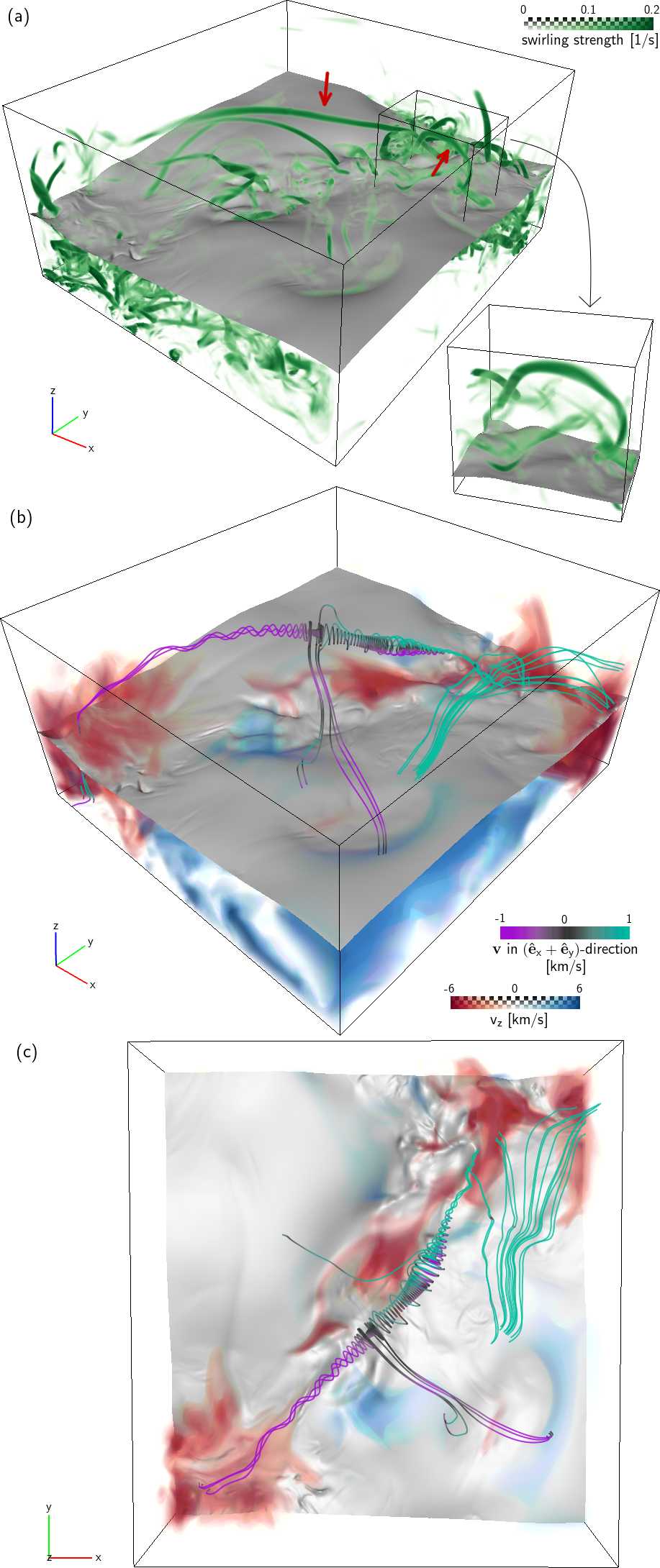}
\caption{Snapshot of Run~H with vortex arcs.  All plots display the warped
\(\tau=1\) surface in gray color.  Panel~(a) shows the swirling strength (green
volume rendering), red arrows indicate the features discussed in the text.
Panels~(b) and~(c) show selected streamlines and the vertical velocity
(red/blue volume rendering). The streamlines are colored according to the
horizontal velocity in the approximate direction of the big arc.  The size of
the box shown is \(1.3\times1.4\times0.7\unit{Mm^3}\). The blow-up
in panel (a) contains the feature presented in Fig.~\ref{fig:bogen2}.}
\label{fig:bogen1}
\end{figure}

\begin{figure}[t]
\centering
\includegraphics[width=\linewidth]{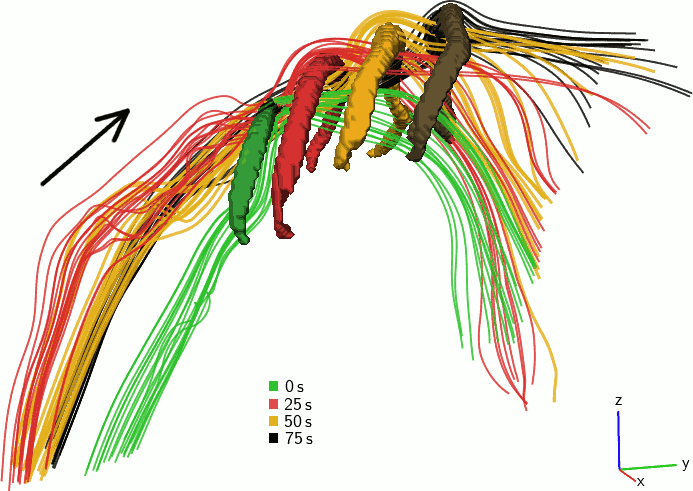}
\caption{Evolution of the small vortex arc which is shown in a blow-up in
Fig.~\ref{fig:bogen1}.  The isosurfaces show the arc at four different times
with a separation of 25 seconds. For each stage, selected velocity streamlines
are plotted in the corresponding color.}
\label{fig:bogen2}
\end{figure}

Fig.~\ref{fig:bogevol} displays the formation of a vortex arc.  When the
feature emerges through the optical surface, it is rather weak and hard to
distinguish from the background. As it rises to its peak height, it gains in
swirling strength.  The horizontal distance between the two footpoints on the
optical surface is \(\simm780\unit{km}\) at this stage, the height above the
surface is \(\simm230\unit{km}\). At last, the feature develops a kink near its
crest and collapses. In total, it lives for approximately 4 minutes.

Fig.~\ref{fig:bogen1} depicts a section from a snapshot of Run~H which contains
a big, slowly evolving vortex arc that hovers high above the optical surface.
Next to it, on the right-hand side in the plots, is a much smaller and rapidly
moving vortex arc.  The two arcs are indicated with red arrows in panel~(a).
In the case of the big arc, the separation of the footpoints of the streamlines
on the optical surface is \(\simm1.4\unit{Mm}\), the peak height above the
optical surface is \(\simm330\unit{km}\).  The vortical fluid motion, indicated
by the large swirling strength in panel~(a), is well visible in the streamlines
plotted in panels~(b) and~(c).  The purple and turquoise colors indicate the
horizontal direction of the fluid motion: material flows from outside towards
the big arc's crest and down on both of its legs.  At the part of the big arc
where the streamlines are purple on one side and turquoise on the other, it is
subject to longitudinal shear.

The small vortex arc has a horizontal extent of \(\simm310\unit{km}\) and a
peak height of \(\simm210\unit{km}\) above the optical surface.  Moving rapidly
with respect to the computational frame of reference, it does not show vortical
streamlines in panels~(b) and~(c) of Fig.~\ref{fig:bogen1}.  Its evolution is
depicted in Fig.~\ref{fig:bogen2}. In \(75\unit{s}\), it propagates
\(\simm200\unit{km}\) horizontally and rises \(\simm100\unit{km}\) vertically.
The corresponding velocities are \(2.7\unit{km\,s^{-1}}\) and
\(1.3\unit{km\,s^{-1}}\), respectively.  The streamlines show that it
corresponds to the front of an upward flow.  In the upper, horizontal part of
the arc, the mean density is reduced to 87\% of the overall horizontal mean.  A
crude estimate of the buoyant rise velocity limited by aerodynamic drag
\citep[Eq.~6 in][]{1975Parker} yields \(\mathord{\sim}1.6\unit{km\,s^{-1}}\).
While this is consistent with the measured vertical velocity, it is not clear
whether buoyancy plays an important active role in the vertical motion
\citep[cf.][]{1993Arendt}.

\subsection{Relationship with larger-scale vortical motion}

\begin{figure*}[t]
\includegraphics[width=\linewidth]{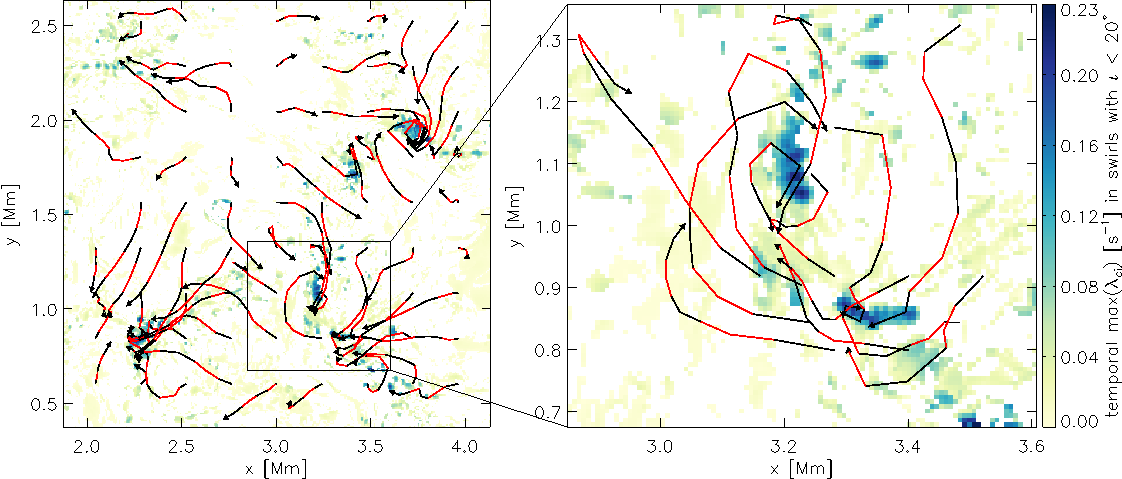}
\caption{Horizontal pathlines in a 177 second time interval, plotted on top of
a map of the temporal maximum of the swirling strength in vertically oriented
swirls, at the average height of the optical surface. Each segment (black or
red) of a pathline corresponds to 35 seconds.}
\label{fig:cork}
\end{figure*}

As described in the introduction, observational evidence for vortical motion on
the solar surface typically refers to larger scales than the strong,
small-scale vortices studied here, although the general characteristics are
similar (e.g., the association with downflows).  The reported mean lifetimes of
granular-scale vortices of 5--8 minutes \citep{2008Bonet,2010Bonet} are not
drastically different from the value of $3.5\unit{min}$ estimated above.
However, because their rotation periods are different, our vortices make about
2 revolutions during their lifetimes, while the observed vortices can be
followed for only a fraction of one rotation \citep[for instance, about 25\% of
a period in the case of][]{2010Bonet}.  It is conceivable that the observed
vortical motions represent the peripheral parts of the much stronger
small-scale vortex cores that show up in the simulations but are too small to
be observed directly. The outer vortex parts would be much more strongly
affected by the evolving granulation pattern and thus be detectable only for a
fraction of a rotation period.

To see whether the vortices studied in this paper would be detectable
in observational data through feature tracking techniques, we consider
horizontal pseudo pathlines (trajectories of fluid elements) in
Fig.~\ref{fig:cork}.  The pathlines are determined from 10 snapshots
of the horizontal velocity field at the average height of the optical
surface with a temporal spacing of \(\mathord{\approx}18\unit{s}\).
They are ``pseudo'' because the vertical velocity is ignored.

Some pathlines are spiraling in towards a region of high swirling strength, see
the feature at \((x,y)\approx(3.7,1.9)\unit{Mm}\) in the left-hand panel of
Fig.~\ref{fig:cork}.  In this case, there is a clear association with the
small-scale vortices presented in this paper.  However, we also find pathlines
which are curved but for which the association with an actual vortex is not
clear, for instance in the blown-up region shown in Fig.~\ref{fig:cork}.

\section{Summary and discussion}
\label{sec:discussion}

We have investigated vortical fluid motions in simulations of near-surface
solar convection by calculating the eigenvalues and eigenvectors of the
velocity gradient tensor field.  Complex eigenvalues with a large imaginary
part indicate regions of strong swirling.  They are found predominantly in and
near the intergranular lanes, where cooled fluid is sinking down in a turbulent
fashion.  The swirling regions form an unsteady network of highly tangled
filaments, some of which protrude above the optical surface.

Near the optical surface, vertically oriented swirls are preferentially located
in the interior of intergranular lanes, where the downflow is strong.
Horizontal swirls, on the other hand, are predominantly located at the edges of
the granules, where vertical motion is mostly absent.  The 3D structure of
contiguous features above the optical surface is manifold, but often in the
form of bent and arc-shaped filaments. These type of structures have previously
been seen in independent numerical simulations with different codes, notably
\citet{2010Muthsam} and \citet{1998Stein}. The swirling direction (rotation
axis) is typically aligned with the longitudinal direction(s) in a contiguous
feature.

Rotary motion causes dynamic pressure by centrifugal forces. One may,
therefore, expect that vortex tubes are underdense as a result of pressure
equilibrium and thermalization with the environment, analogous to magnetic flux
tubes.  We find vortex features above the optical surface to be underdense by
\(\simm15\%\) with respect to the horizontal mean in downflow regions. A
significant depression of the optical surface is common where it intersects
with vortices.  Having diameters of \(\simm80\unit{km}\), the largest vertical
vortex features would be visible as bright spots with a size of \(\simm0.1\)
arcseconds in observations of the Sun.

Most vortex features are highly unsteady, being moved and/or twisted by
turbulent fluid motions.  Some of them rise upwards.  Although, in principle,
buoyancy could contribute to this rise (\citealt{1991Parker}, but see also
\citealt{1993Arendt}) we find no unambiguous evidence for the operation of
buoyancy forces.  After a typical lifetime of a few minutes, the features
usually descend below the optical surface, where they dissolve.

In general, the vortices presented here are smaller than those found in
observations so far. They may represent peripheral parts of strong, small-scale
vortices. However, arc-like motions of surface features do not necessarily
imply the presence of a strong central vortex.  The pattern of the fluid flow
is changing considerably at time scales of only a few minutes, which is
reflected by the lifetimes of the small-scale vortices.  We therefore reckon
that granular-scale or bigger vortical flows are less numerous than small-scale
vortices.

\begin{acknowledgements}
This work has been supported by the Max-Planck Society in the
framework of the Interinstitutional Research Initiative \textit{Turbulent
transport and ion heating, reconnection and electron acceleration in
solar and fusion plasmas} of the MPI for Solar System Research,
Katlenburg-Lindau, and the Institute for Plasma Physics, Garching
(project MIF-IF-A-AERO8047).
\end{acknowledgements}

\bibliography{ref}

\end{document}